\documentclass[11pt,a4paper]{article}
\pdfoutput=1
\usepackage{jheppub}

\usepackage{amsmath}
\usepackage{bm}
\usepackage{braket}
\usepackage{graphicx}
\usepackage{tikz}
\usetikzlibrary{matrix}
\usepackage{multirow}

\newcommand{\Oi}{\mathcal{O}}
\newcommand{\be}{\begin{equation}}
\newcommand{\ee}{\end{equation}}

\newcommand{\tsigma}{\tilde{\sigma}}

\subheader{\begin{flushright} UTTG-32-13, TCC-026-13 \end{flushright}}

\title{Decelerating cosmologies are de-scramblers}
\author{Daniel Carney, Willy Fischler}

\affiliation{Theory Group, Department of Physics, University of Texas at Austin and \\ Texas Cosmology Center \\ Austin, Texas 78712}

\emailAdd{carney@physics.utexas.edu}
\emailAdd{fischler@physics.utexas.edu}

\abstract{Stationary observers in static spacetimes see falling objects spread exponentially fast, or fast-scramble, near event horizons. We generalize this picture to arbitrary cosmological horizons. We give examples of exponential fast-scrambling and power-law scrambling and ``de-scrambling'' as charges propagate freely near a horizon. In particular we show that when the universe is decelerating, information hidden behind the apparent horizon is de-scrambled as it re-enters the view of the observer. In contrast to the de Sitter case, the power-law scaling suggests that the microscopic dynamics of the horizon are local.}

\arxivnumber{13xx.yyyy}

\begin{document}
\unitlength = 1mm

\maketitle

\section{Introduction}
Data strongly indicates that our universe has large-scale causal horizons.\cite{Riess:1998cb,Ade:2013zuv} In particular, if the scale factor of the universe is accelerating in the asymptotic future, our observations are bounded by a cosmological event horizon. Thus we have causal access to only a portion of the universe.

Physics in the presence of horizons is subtle.\cite{Hawking:1974sw,Gibbons:1977mu} One must be careful to distinguish statements with operational meaning in terms of physical observation, a point strongly expressed by the idea of complimentarity.\cite{thooft,Susskind:1993if} In this paper we formulate physics in FRW spacetimes in the reference frame of a physical observer, a coordinate chart covering precisely the events in his past lightcone.\cite{Klein:2010rk} We show how to map calculations in co-moving coordinates into the frame.

In searching for a quantum formulation of physics in cosmological spacetimes, a natural question to ask is what happens to localized information as it nears the edge of observational range. The answer is known when spacetime is static: the observer sees localized information like a charge or string spread exponentially fast or ``fast-scramble'' across the horizon.\cite{Thorne:1986iy,Sekino:2008he,Susskind:2011ap} In the context of holography, an equivalent statement should hold in the dual. Locally interacting degrees of freedom are only known to be capable of spreading information at power-law rates, so this strongly suggests that the scrambling of information on the horizon is controlled by non-local processes at the microscopic level.\cite{Susskind:2011ap}

The horizons present in nature are not static: black holes accrete and evaporate, and the cosmological horizons move in response to the expansion of the universe. In this paper, we consider the same question for horizons which move in time. We calculate the scrambling rates for point charges propagating freely near arbitrary cosmological horizons.\footnote{In particular, we assume that the dynamics remains free even when the charge is behind a horizon.} We show that information is scrambled onto the apparent horizon of a co-moving observer if the scale factor is accelerating. Conversely, when the universe is decelerating, the observer sees the charge ``de-scramble'' as it re-enters the horizon. The rate of (de-)scrambling is slower than exponential except during de Sitter-like expansion. This suggests that one may be able to describe the observations of a physical observer in terms of locally interacting degrees of freedom.

The paper is organized as follows. We state the problem precisely and review what is known in section \ref{problemsection}. In section \ref{framesection} we construct the observer's frame of reference, and develop the coordinate transformation from co-moving coordinates. Section \ref{scramblingsection} exhibits the scrambling rates of an inertial charge falling near the apparent horizon; we give the general answer and study some physically relevant examples. We state our conclusions in section \ref{conclusionsection}. For the convenience of the reader we also review the causal structure of flat FRW universes (in co-moving coordinates) in appendix \ref{causalstructureappendix}.

\begin{figure}[h]
\centering
\includegraphics[scale=0.5]{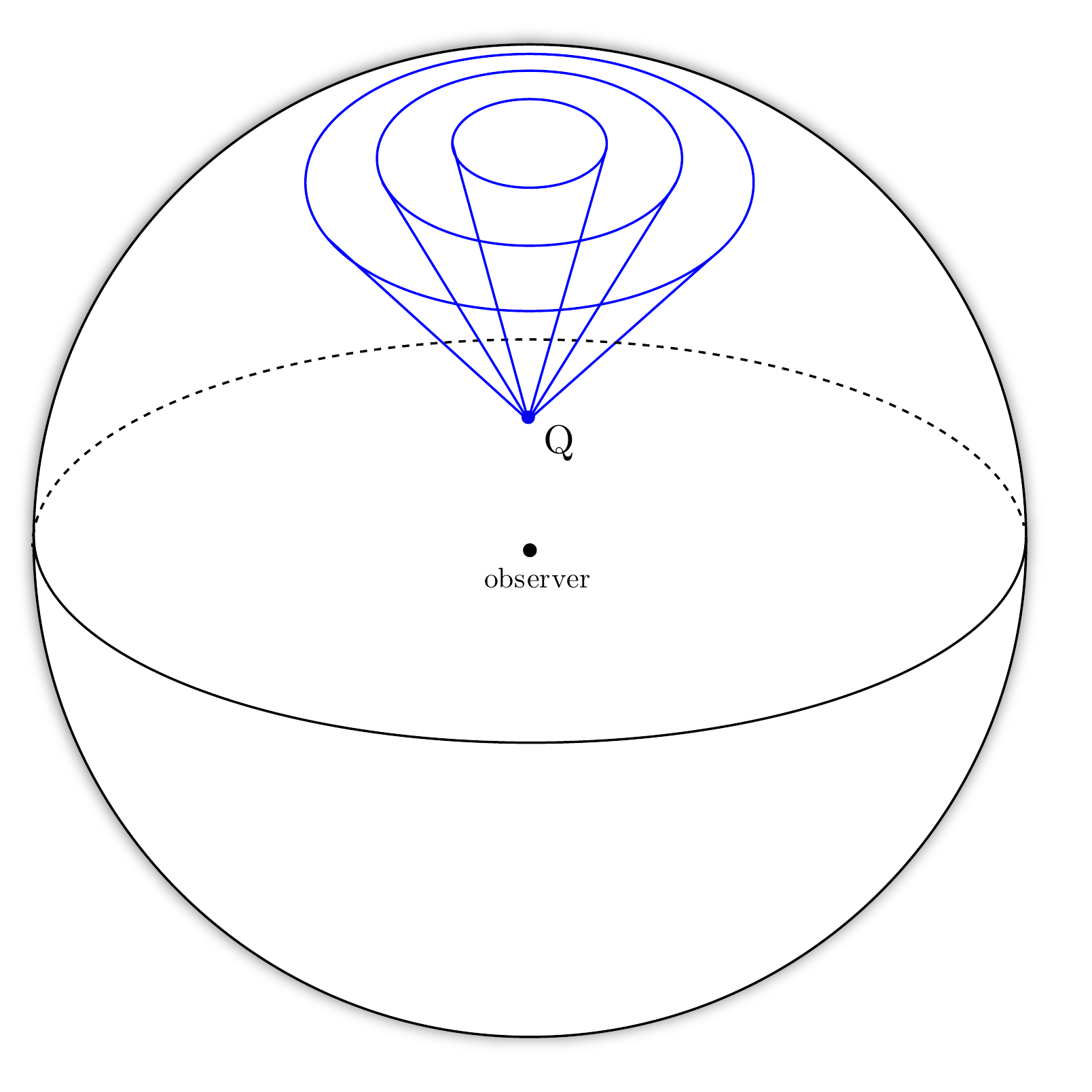} \\
\caption{Point charge $Q$ projecting its image onto the horizon of a co-moving observer $\Oi$. This picture represents the situation on a spatial slice in the observer's frame at some fixed observer time $\tau$ (section \ref{framesection}).}
\label{spherepicture}
\end{figure}

\section{Review and statement of the problem}
\label{problemsection}

Scrambling is the process by which some localized information spreads out into a larger system. Early studies (eg. \cite{Thorne:1986iy,Sekino:2008he}) considered the example of the electric field of a point charge falling onto a static event horizon. It was found that the electric field of the charge spreads exponentially fast across a stretched horizon placed a very short distance away from the event horizon. Later it was pointed out that the same conclusions held for charges freely propagating in the near-horizon region of de Sitter space.\cite{Susskind:2011ap}

The scrambling process is intimately tied up with unitarity. One of the original motivations for its study was to understand how a quantum mechanical system obeying unitarity can ``thermalize'' a local perturbation.\cite{vonneumann} Quantum mechanically, we say that a small subsystem of a system in a pure state is scrambled as the subsystem becomes entangled with the rest of the system. In a local quantum field theory, this process occurs at a power-law rate in time. The simplest example to understand is diffusion, which in $d$ dimensions gives a scrambling rate $\sim t^{d/2}$. 

Susskind and Sekino therefore conjectured in \cite{Sekino:2008he} that the exponential ``fast scrambling'' on the horizon is controlled by non-local processes at the microscopic level, for example some kind of matrix quantum mechanics. An explicit quantum fast-scrambler was constructed as a quantum circuit in \cite{dankert}; attempts at more physical examples can be found in eg. \cite{Edalati:2012jj,Lashkari:2011yi}.

We do not currently have an acceptable unitary quantum theory in the presence of causal horizons, except in anti-de Sitter space\cite{Maldacena:1997re} and in some flat backgrounds.\cite{Banks:1996vh} In cosmological spacetimes the situation is worse: even the correct choice of observables is unclear (see eg. \cite{Banks:2001yp,Witten:2001kn}). In the semi-classical approach typically used, the observables are taken to be correlation functions of field operators, say on a spacelike slice $\Sigma$ in co-moving coordinates ${ds^2 = -dt^2 + a^2(t) d\Sigma^2}$. Depending on the scale factor $a(t)$ and spatial curvature of $\Sigma$, a physical observer may not have causal access to all of $\Sigma$. Therefore the operational meaning of unitarity in such a formulation is unclear.\footnote{Of course, one can implement time evolution through a unitary operator $U$ generating dynamics between slices.\cite{Weinberg:2005vy} But this is pure mathematics: physically, unitarity is the requirement that the sum of the probabilities of all possible outcomes of any given measurement is unity, and no observer can measure all of these correlation functions. (See eg. \cite{Bousso:2011up} for an alternative view). The only reason one can get away with this in practice is due to a lucky separation of scales between the inflationary and modern Hubble parameters, $H_{inf}/H_0 \gtrsim 10^{40}$, which means that to very good approximation we can view ourselves as a ``meta-observer'' who can see all the modes relevant to the early inflationary period.\cite{Anninos:2012qw}}

The purpose of this paper is to take steps towards a more physical formulation of observation in cosmological spacetimes. We focus on the scrambling of point electric charges for definiteness. The most straightforward approach to unitarity would be to only consider observables which are causally connected to a physical observer. In order to implement unitary in this sense, we show how to use recent results of Klein and Randles \cite{Klein:2010rk,Klein:2012cp} to construct a coordinate system which covers precisely the set of events that can be observed in principle by an (immortal) inertial observer in an arbitrary FRW spacetime. We show how to map tensorial quantities (like the field strength of a charge) known in FRW coordinates into the observer's frame. Thus we can adapt the successful calculations already understood in co-moving coordinates and study them in terms of physical observations without reference to causally disconnected regions of spacetime.

We ask what a co-moving observer sees as a point charge propagates freely near the apparent horizon. The answer is simple and intuitive. We find that if the scale factor of the universe is accelerating, the apparent horizon acts as a scrambler as the charge passes through the horizon. The scrambling rate is exponential for exponential inflation (recovering the result of \cite{Susskind:2011ap}), but only power-law in general. Conversely if the universe is decelerating, the images of charges which were previously scrambled across the horizon will appear to coalesce back into a point followed by the charge re-appearing inside the horizon. We call this phenomenon ``de-scrambling''; it also generically occurs at power-law rates. Put together, these results suggest that one may be able to model the ``causal patch'' of a physical observer in a cosmological spacetime using locally interacting degrees of freedom. 

\section{Observer frame}
\label{framesection}

In order to study things as seen by a physical observer, we will construct coordinates based on a non-rotating frame of reference centered on his worldline.\cite{fermi,walker,Misner:1974qy,Poisson:2011nh,Klein:2010rk,Klein:2012cp} These coordinates offer a number of technical advantages over co-moving coordinates. In particular, for a co-moving observer in FRW spacetimes, Klein\cite{Klein:2012cp} has proven that these coordinates cover precisely the causal past of the observer at future infinity, i.e. his past lightcone, which makes them ideally suited to our purpose. Many quantities, like the metric or any other tensorial quantity known in FRW coordinates, turn out to be reasonably simple to express.\footnote{Section \ref{geometrysection} follows \cite{Klein:2010rk,Klein:2012cp} closely; we thank David Klein for discussions on those papers.}

\subsection{Geometry}
\label{geometrysection}
Consider some fixed cosmological spacetime described by a flat Friedmann-Robertson-Walker metric,
\be
\label{frwmetric}
ds^2 = -dt^2 + a^2(t) dr^2 + a^2(t) r^2 d\Omega^2.
\ee
In what follows we assume the scale factor $a(t)$ is smooth, monotonic, non-decreasing and we will typically assume it asymptotes to $a(t_0)=0$ at the beginning of time, at redshift $z \rightarrow \infty$.\footnote{For a big bang cosmology this means the big bang hypersurface $t=t_0$ (we often take $t_0=0$). We also consider cosmologies which are exponentially inflating in the infinite past $t_0 \rightarrow -\infty$. Later we will drop the smoothness assumptions to allow for phase transitions.} In particular we do not need to assume the scale factor solves the Friedmann equations.

To construct the Fermi-Walker coordinates, one considers a timelike observer $\Oi$ defined by his worldline $\Oi(\tau)$, with $\tau$ his proper time. We take spacelike slices at time $\tau$ to be generated by the set of spacelike geodesics orthogonal to $d\Oi/d\tau$. Pick an event $p$. Consider all spacelike geodesics through $p$. If $p$ is sufficiently close to $\Oi$, precisely one such radial geodesic will orthogonally intersect $\Oi$. Let $\tau$ denote the time of this intersection, $\rho$ the proper distance along the geodesic from $\Oi(\tau)$ to $p$, and $(\theta,\phi)$ its angular coordinates. Then we define the frame coordinates or Fermi-Walker coordinates of $p$ to be 
\be
x^{\hat{a}} := (\tau,\rho,\theta,\phi).
\ee

In either FRW or frame coordinates we have a coordinate singularity at $r=0$ where the angular coordinates degenerate; in order to exploit the spherical symmetry of our spacetime what we will do is work out the frame in the $t-r$ plane, and re-instate the angular coordinates at the end. Doing so, we will show that the metric expressed in Fermi-Walker coordinates takes the form\cite{Klein:2010rk}
\be
\label{fwmetric}
ds^2 = -g_{\tau\tau}(\tau,\sigma) d\tau^2 + d\rho^2 + R^2(\tau,\sigma) d\Omega^2,
\ee
where $\sigma = \sigma(\tau,\rho)$ is a function measuring the redshift of the event located at $(\tau,\rho)$ given below, and $R^2 = a^2 r^2$ measures the proper area of the horizon. The metric coefficients have the property that $ds^2$ is just the flat metric along $\Oi$'s worldline. In the rest of this section we derive the metric coefficients; along the way we will work out the transformation rules for arbitrary tensorial quantities.

Clearly our main task is to work out the spacelike geodesics orthogonal to $\Oi$. From here out we take $\Oi$ to be an inertial observer at the spatial origin of co-moving coordinates (\ref{frwmetric}), without loss of generality. Fix a time $\tau$ along the worldline. Denote the geodesic we want by $\gamma(\rho) = (t(\rho),r(\rho))$ where $\rho$ is proper distance along the geodesic; we normalize $\rho=0$ on $\Oi$. Since the geodesics are spacelike they will minimize the proper length
\be
L[\gamma] = \int_{0}^{\rho} d\rho' \left[ - \left( \frac{dt}{d\rho} \right)^2 + a^2(t) \left( \frac{dr}{d\rho} \right)^2 \right]^{-\frac{1}{2}}.
\ee
One immediately sees that $a^2(t) dr/d\rho = C$ is constant along the geodesic. Demanding that $\rho$ is proper length and that the geodesic is normal to $\Oi$ at $\rho=0$ tells us that $C = a(\tau)$ and $dt/d\rho = \pm \sqrt{a^2(\tau)/a^2(t)-1}$. The geodesic minimizes spatial length, and $a(t)$ decreases as $t$ runs back into the past, so we must take the minus sign. 

To integrate the geodesic equation it is convenient to use the parameter
\be
\label{sigmadef}
\sigma = \frac{a^2(\tau)}{a^2(t)} = \left( 1 + z \right)^2.
\ee
Here the second equality points out that $\sigma$ is directly related to the redshift between the event along the geodesic, which has FRW time $t$, and the observer's time $\tau$. Then in all we have $dt/d\rho = - \sqrt{\sigma-1}$. Clearly $\sigma = 1$ when the geodesic originates on $\Oi$'s worldline and increases as $\rho$ increases, and $\sigma \rightarrow \infty$ as the geodesic runs arbitrarily backward in cosmic time $t$.

The geodesics can be written in integral form in terms of $\tau$ and $\sigma$. We can also get a formula for the proper length $\rho$ along the geodesics. These are sufficient to transform any tensor into the frame. Let $b$ denote the inverse of the scale factor, i.e. the function such that $b(a(t)) = t$. Inverting (\ref{sigmadef}) gives the FRW time in terms of observer time $\tau$ and the redshift along the geodesic:
\be
\label{tcoord}
t(\tau,\sigma) = b\left( \frac{a(\tau)}{\sqrt{\sigma}} \right).
\ee
Re-arranging (\ref{sigmadef}) as $a(t) = a(\tau)/\sqrt{\sigma}$, differentiating with respect to $\rho$, and using the inverse function theorem to write $b'(a(t)) = 1/\dot{a}(t)$ one finds
\be
\label{rhocoord}
\rho(\tau,\sigma) = \frac{a(\tau)}{2} \int_{1}^{\sigma} b' \left( \frac{a(\tau)}{\sqrt{\tsigma}} \right) \frac{d\tsigma}{\tsigma^{3/2} \sqrt{\tsigma-1}}.
\ee
To get the co-moving radial coordinate $r=r(\tau,\sigma)$, note that we have
\be
\frac{dr}{d\rho} = \frac{dr}{d\sigma}\frac{d\sigma}{d\rho};
\ee
solving this for $dr/d\sigma$ and using similar manipulations we find
\be
\label{rcoord}
r(\tau,\sigma) = \frac{1}{2} \int_{1}^{\sigma} b' \left( \frac{a(\tau)}{\sqrt{\tsigma}} \right) \frac{d\tsigma}{\tsigma^{1/2} \sqrt{\tsigma-1}}.
\ee

In order to transform co-moving quantities into the frame we need to work out the derivatives of the coordinate transformation. The equations above define a set of coordinate transformations between coordinates $\{t,r\}$, $\{\tau,\sigma\}$, and $\{\tau,\rho\}$. The situation is summarized by the diagram:
\be 
\label{coorddiagram}
\begin{tikzpicture}[every node/.style={midway}]
\matrix[column sep={6em,between origins},
        row sep={4em}] at (0,0)
{ \node(U)   {$\{ \tau,\sigma \}$}  ; & \node(V) {$\{ t,r \}= y^{\mu}$}; \\
  \node(W) {$x^{\hat{a}} = \{ \tau,\rho \}$};                   \\};
\draw[<-] (W) -- (U) node[anchor=east]  {$G$};
\draw[<-] (W) -- (V) node[anchor=north]  {$H$};
\draw[->] (U)   -- (V) node[anchor=south] {$F$};
\end{tikzpicture}
\ee
where the images are given by (\ref{tcoord}), (\ref{rhocoord}), (\ref{rcoord}), and composition. The $\{\tau,\sigma\}$ coordinates express the geometry in terms of redshifts directly, but lead to messy formulas (in particular a non-diagonal metric). The transformation to Fermi-Walker coordinates, in which the metric takes the form (\ref{fwmetric}), is given by the map $H = G \circ F^{-1}$. To transform covariant tensors we also want the derivatives of this map. Doing some calculus with (\ref{coorddiagram}) one finds that
\be
\Lambda^{\mu}_{\ \hat{a}} = (dH^{-1})^{\mu}_{\ \hat{a}} = \begin{pmatrix} \Lambda^{t}_{\ \tau} & \Lambda^{t}_{\ \rho} \\ \Lambda^{r}_{\ \tau} & \Lambda^{r}_{\ \rho} \end{pmatrix}
\ee
where the coefficients are, after some integrations by parts,
\be
\label{lambdacomponents}
\begin{split}
\Lambda^{t}_{\ \tau} & = \frac{\partial t}{\partial \tau} - \frac{\partial \rho}{\partial \tau} \frac{\partial t/\partial \sigma}{\partial \rho/\partial \sigma} = \dot{a}(\tau) \sqrt{\sigma} \mathcal{F}(\tau,\sigma) \\
\Lambda^{r}_{\ \tau} & = \frac{\partial r}{\partial \tau} - \frac{\partial \rho}{\partial \tau} \frac{\partial r/\partial \sigma}{\partial \rho/\partial \sigma} = -\frac{\dot{a}(\tau)}{a(\tau)} \mathcal{F}(\tau,\sigma) \sqrt{\sigma(\sigma-1)} \\
\Lambda^{t}_{\ \rho} & = \frac{\partial t/\partial \sigma}{\partial \rho/\partial \sigma} = -\sqrt{\sigma-1} \\
\Lambda^{r}_{\ \rho} & = \frac{\partial r/\partial \sigma}{\partial \rho/\partial \sigma} = \frac{\sigma}{a(\tau)}.
\end{split}
\ee
In these formulas, the function $\mathcal{F}$ is given by
\be
\mathcal{F}(\tau,\sigma) = \left[ b' \left( \frac{a(\tau)}{\sqrt{\sigma}} \right) + a(\tau) \mathcal{I}(\tau,\sigma) \sqrt{\frac{\sigma - 1}{\sigma}} \right],
\ee
where $\mathcal{I}$ is the integral
\be
\label{integral}
\mathcal{I} = \mathcal{I}(\tau,\sigma) = \frac{1}{2} \int_{1}^{\sigma} b'' \left( \frac{a(\tau)}{\sqrt{\tilde{\sigma}}} \right) \frac{d\tilde{\sigma}}{\tilde{\sigma}\sqrt{\tilde{\sigma}-1}}.
\ee
To do the full four-dimensional transformations one just maps the angular coordinates with the identity, i.e. $\Lambda_{\theta}^{\theta} = \Lambda_{\phi}^{\phi} = 1$, with all other components vanishing.

With these expressions in hand, we are ready to work out any tensorial quantities in the frame. As a warmup it is a good exercise to check that the metric transforms correctly to the Fermi-Walker form (\ref{fwmetric}). Transforming from FRW coordinates $g_{\hat{a}\hat{b}} = \Lambda^{\mu}_{\hat{a}} \Lambda^{\nu}_{\hat{b}} g_{\mu\nu}$ and writing $a(t)$ using (\ref{sigmadef}) one finds that the $\rho-\rho$ component is
\be
g_{\rho\rho}  = 1.
\ee
Similar but slightly more involved manipulations give
\be
\label{gtautau}
g_{\tau\tau} = -\dot{a}^2(\tau) \mathcal{F}^2(\tau,\sigma).
\ee
The metric components along the spheres also transform: we get
\be
\label{gsphere}
g_{\theta\theta} = R^2(\tau,\sigma) := a^2(\tau) r^2(\tau,\sigma)/\sigma, \ \ g_{\phi\phi} = R^2(\tau,\sigma) \sin^2 \theta.
\ee
It is straightforward to show by direct calculation that the off-diagonal metric coefficients vanish. These results reproduce those in \cite{Klein:2010rk}.

\subsection{Causal structure}
\label{causalstructuresection}

\begin{figure}[h]
\centering
\includegraphics[scale=0.55]{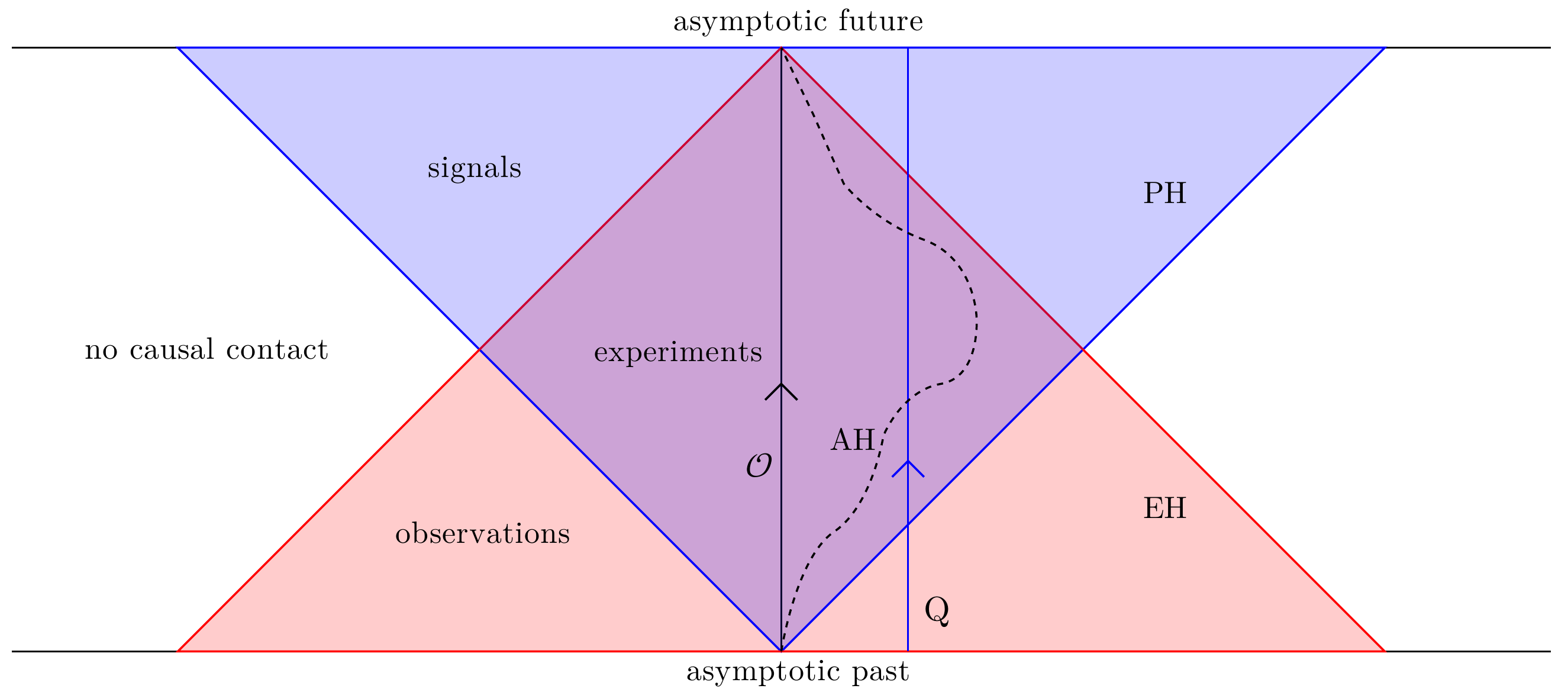} \\
\caption{Causal structure of a flat FRW universe, according to an inertial (co-moving) observer $\Oi$. For simplicity we have plotted the situation in conformal co-moving coordinates ${ds^2 = a^2(\eta) (-d\eta^2 + dr^2 + r^2 d\Omega^2)}$, with the past $\eta \rightarrow -\infty$ compactified in the inflationary case. Time runs from bottom to top, the radial coordinate increases outward from the observer, and we have suppressed the spacelike $S^2$ at each point.}
\label{causalstructurepicture}
\end{figure}

Observers in a cosmological spacetime may experience event, particle, and/or apparent horizons. The first two are global structures depending on the full history of the universe and define the boundaries of causal accessibility of the observer, see figure \ref{causalstructurepicture}. The apparent horizon, on the other hand, is locally defined in time.\footnote{The co-moving radius of the apparent horizon is the scale that enters the equations of motion of cosmological perturbations, eg. modes of co-moving wavenumber $k$ ``cross the horizon'' when $k r_{AH} = 1$.} The crucial difference for us is that the global horizons are null hypersurfaces whereas the apparent horizon can take any spacetime signature.

We review the causal structure of flat FRW geometries in co-moving coordinates in some detail in appendix \ref{causalstructureappendix}. Here we record the basic results on the causal structure that we need. The event horizon of an inertial observer $\Oi$ exists if and only if the integral $r_{EH}(t) = \int_t^{\infty} dt' / a(t')$ converges, see eq. (\ref{eventhorizonco-moving}), in other words if the universe is accelerating in the asymptotic future. The observer's event horizon is the hypersurface consisting of spheres along this curve in the $t-r$ plane. The existence and size of the event horizon at time $t$ depend on the whole history of the universe in the future of $t$. As demonstrated in \cite{Klein:2012cp}, the frame coordinates are bounded by the event horizon; if $\Oi$ does not have one then his frame covers the entire cosmology. One can locate the frame explicitly by transforming $r_{EH}(t)$ with (\ref{tcoord}), (\ref{rcoord}); if it exists it is at infinite redshift $\sigma \rightarrow \infty$.

The apparent horizon of $\Oi$ is determined locally in time. At any given time, lightrays orthogonal to the sphere at a certain radius may appear to $\Oi$ to be stuck in time. That is, at least one of the families of null geodesics orthogonal to the spheres will have vanishing geodesic expansion, see appendix \ref{causalstructureappendix}. In co-moving coordinates these spheres lie along the line $r_{AH}(t) = 1/a(t) H(t)$ and form the apparent horizon. To find the apparent horizon in the frame, the easiest thing to do is notice that the condition that the expansion of the null congruence with tangent field $k$ vanishes, $\Theta = \nabla_k k = 0$, is invariant under transformations on the $t-r$ plane. Thus we can locate the apparent horizon by solving
\be
\label{easyhorizon}
\begin{pmatrix} t \\ r_{AH}(t) \end{pmatrix} = \begin{pmatrix} t(\tau,\sigma_{AH}(\tau)) \\ r(\tau,\sigma_{AH}(\tau)) \end{pmatrix}
\ee
to find the redshift parameter $\sigma_{AH}(\tau)$ at which the horizon resides. The apparent horizon need not be generated by null geodesics. If the spacetime is static then the apparent horizon coincides with the event horizon and is null. Generically however, the apparent horizon can take any spacetime signature; in particular it is spacelike if the scale factor is decelerating. It is this fact that allows us to make sense of charges passing through the apparent horizon.

\section{(De-)scrambling}
\label{scramblingsection}
Scrambling is the propagation of local information, say a perturbation around some state, as it interacts with the system at large. Holography suggests that we study the bulk physics in terms of some kind of projection onto the horizon. The simplest classical implementation is to note that, given the history of the universe and the Maxwell equations, the worldline of a point charge and the electric field it induces on the horizon are equivalent pieces of data.\footnote{We are neglecting any interactions, in particular the backreaction of the charge and its field.} In other words, we can trade the boundary condition for the solution at the classical level. One can make it even more clear by defining the induced surface charge on the horizon: then scrambling is the statement that the induced charge density spreads out in time as the charge nears the horizon.

In various static black holes and in de Sitter spacetime it is known that the charge induced by a geodesic point charge $Q$ near the event horizon spreads exponentially fast over the stretched horizon at a timescale $M_{BH}^{-1}$ or $H^{-1}$, respectively.\cite{} In both cases this can be understood by considering a near-horizon limit given by Rindler coordinates.\cite{susskindbook} In the Rindler case the timescale is the inverse of the proper acceleration required to keep an observer $\Oi$ at fixed radial distance $\epsilon$ from the horizon, $a \sim 1/\epsilon$; fortunately Rindler space has a scaling symmetry and one can send $\epsilon \rightarrow H^{-1}$ at the end. In all cases the calculations are simple because the metric is time-independent. Nonetheless this is suggestive that one should consider the dynamics as viewed by some particular observer $\Oi$.

What we will do is calculate the angular distribution of charged induced on the apparent horizon of a co-moving observer $\Oi$ watching a point charge $Q$ falling near the horizon. One can interpret the calculations in a simple way, following \cite{Sekino:2008he,Susskind:2011ap}. While the charge $Q$ is inside the horizon, it is just bulk data satisfying the Maxwell equations. While the charge is behind the horizon, we instead think of the induced charge $\Sigma$ as the holographic representation of the information.

We define the induced charge using Gauss' law. Suppose we know the electromagnetic field strength $F_{\hat{a}\hat{b}}$ everywhere in the observer's frame. Now consider a small area $dA = R^2(\tau,\sigma_{AH}) \sin \theta d\theta \wedge d\phi$ on the horizon at time $\tau$. The Gauss law $d \star F = \star J$ says that the induced charge $Q_{ind}$ on this area of the horizon is given by
\be
\Sigma(\tau,\theta,\phi) dA = (\star F)_{\theta\phi} (\tau,\sigma_{AH},\theta,\phi) d\theta \wedge d\phi.
\ee
In terms of the radial electric field we have
\be
(\star F)_{\theta\phi} = \sqrt{-g} \epsilon_{\tau\rho\theta\phi} F^{\tau\rho} = \sqrt{-g_{\tau\tau}} R^2 \sin \theta F^{\tau\rho}
\ee
so we identify the surface charge density on the appropriate horizon
\be
\label{chargedensity}
\Sigma = \frac{F_{\tau\rho}}{\sqrt{-g_{\tau\tau}}} \Bigg|_{horizon} = -\frac{Q}{4\pi} \frac{\sigma}{a^2(\tau)} \frac{r(\tau,\sigma) - r_Q \cos \theta}{\Delta r^3(\tau,\sigma)} \Bigg|_{horizon}.
\ee
In evaluating this, one can use either the redshift parameter $\sigma=\sigma_{horizon}(\tau)$ or the radial frame distance $\rho=\rho_{horizon}(\tau)$ of the horizon. Note that in deriving this formula, we are only considering the electric flux on one side of the horizon, i.e. the side facing the observer. 

If the metric is static then the horizon is both an event horizon and apparent horizon. It is a null hypersurface and one finds that the induced charge is just a constant $\Sigma \equiv Q/4\pi H^{-2}$ across the sphere, at any time. In accordance with the membrane paradigm, in this case we can regulate the calculations by looking at the stretched horizon, a timelike hypersurface placed a small frame distance $\rho_{SH} = \rho_{AH,EH} - \epsilon$ from the causal horizon.\cite{Thorne:1986iy,Susskind:1993if}

\subsection{Co-moving electrostatics in the frame}
To study the scrambling of a geodesic point charge we need its field strength. The easiest way to get it is to write down the answer in co-moving coordinates and then transform it to the frame.

Consider a point electric charge $Q$ in an FRW universe. Suppose the charge is co-moving with an inertial observer $\Oi$ at the origin, so it lives on the timelike geodesic $(t,r,\theta,\phi) \equiv (t,r_Q, \theta_Q, \phi_Q)$. If the charge is at the spatial origin, it produces the Coulomb field $F = -Q/4\pi a r^2 dt \wedge dr$.\footnote{The factor of $a$ is fixed by the Gauss law on some co-moving sphere $Q = \int_{S^2} \star F$.} If the charge is displaced from $\Oi$, we can translate this to obtain
\be
F = F_{tr} dt \wedge dr + F_{t\theta} dt \wedge d\theta,
\ee
with, taking the charge along the $z$-axis ($\theta_Q = 0$) for simplicity,
\be
F_{tr} = -\frac{Q}{4 \pi a(t)} \frac{r-r_Q \cos \theta}{\Delta r^3}, \ \  F_{t \theta} = -\frac{Q}{4 \pi a(t)} \frac{r r_Q \sin \theta}{\Delta r^3}.
\ee
Here $\Delta r$ is the co-moving distance from the charge to the spatial origin
\be
\label{deltar}
\Delta r^2 = r^2 - 2 r r_Q \cos \theta + r_Q^2.
\ee

In this expression for $F$ we see a simple way in which FRW coordinates are not so great for describing the observations of $\Oi$: an inertial observer in an expanding universe would see $Q$ receding from view, and thus a current, and so he should see a magnetic field. But this is nowhere to be found in co-moving coordinates, which are defined by the statement that he and the charge have fixed coordinate distance. When we go to our frame coordinates we will see the magnetic field show up again.

Transforming this expression to the frame is straightforward. The components transform as usual $F_{\hat{a}\hat{b}}(x^{\hat{a}}) = \Lambda^{\mu}_{\hat{a}} \Lambda^{\nu}_{\hat{b}} F_{\mu\nu} (y^{\mu}(x^{\hat{a}}))$. One finds after routine computation using (\ref{lambdacomponents}) that we have a radial electric field 
\be
\label{framefield-transverse}
F_{\tau\rho} = F_{\tau\rho}(\tau,\sigma) = -\frac{Q}{4\pi} \frac{\sigma H(\tau) F(\tau,\sigma)}{a(\tau)} \frac{r(\tau,\sigma) - r_Q \cos \theta}{\Delta r^3(\tau,\sigma)}
\ee
where the co-moving radial coordinate is expressed in frame coordinates via (\ref{rcoord}). Here we defined the Hubble rate at frame time $\tau$ as $H(\tau) = \dot{a}(\tau)/a(\tau)$. We also find an electric field tangential to the spatial spheres
\be
F_{\tau\theta} = -\frac{Q}{4\pi} \sigma H(\tau) F(\tau,\sigma) \frac{r(\tau,\sigma) r_Q \sin \theta}{\Delta r^3(\tau,\sigma)}
\ee
and a magnetic field along the azimuth $\phi$,
\be
F_{\rho\theta} = \frac{Q}{4\pi} \frac{\sqrt{\sigma (\sigma-1)}}{a(\tau)} \frac{r(\tau,\sigma) r_Q \sin \theta}{\Delta r^3(\tau,\sigma)}.
\ee
The parameter $r_Q$ represents the initial condition for the charge in co-moving coordinates. We can re-interpret it in the frame as the redshift $\sigma_Q$ which $\Oi$ assigns to $r=r_Q$ at some reference time $\tau=\tau_0$. That is, $\sigma_Q$ is defined by (\ref{rcoord}) as
\be
\label{rqframe}
r_Q = r(\tau_0,\sigma_Q).
\ee

From these expressions, one can see the general behavior of the angular distribution of the induced charge (\ref{chargedensity}). Using (\ref{rcoord}), (\ref{deltar}) and (\ref{rqframe}), we see that the angular dependence is varying in time according to $b'(a(\tau)) \sim e^{-H_0 \tau}$ during exponential inflation or $b'(a(\tau)) \sim \tau^{1-\alpha}$ for a power law $a(\tau) \sim \tau^{\alpha}$. Clearly the behavior for a decelerating cosmology $\alpha < 1$ is opposite that of an accelerating cosmology: accelerating epochs scramble information across the horizon, and decelerating epochs de-scramble it back together. We now turn to some physically relevant examples.

\subsection{Cosmological constant ($w=-1$)}

\begin{figure}[h]
\begin{center}$
\begin{array}{cc}
\includegraphics[scale=0.7]{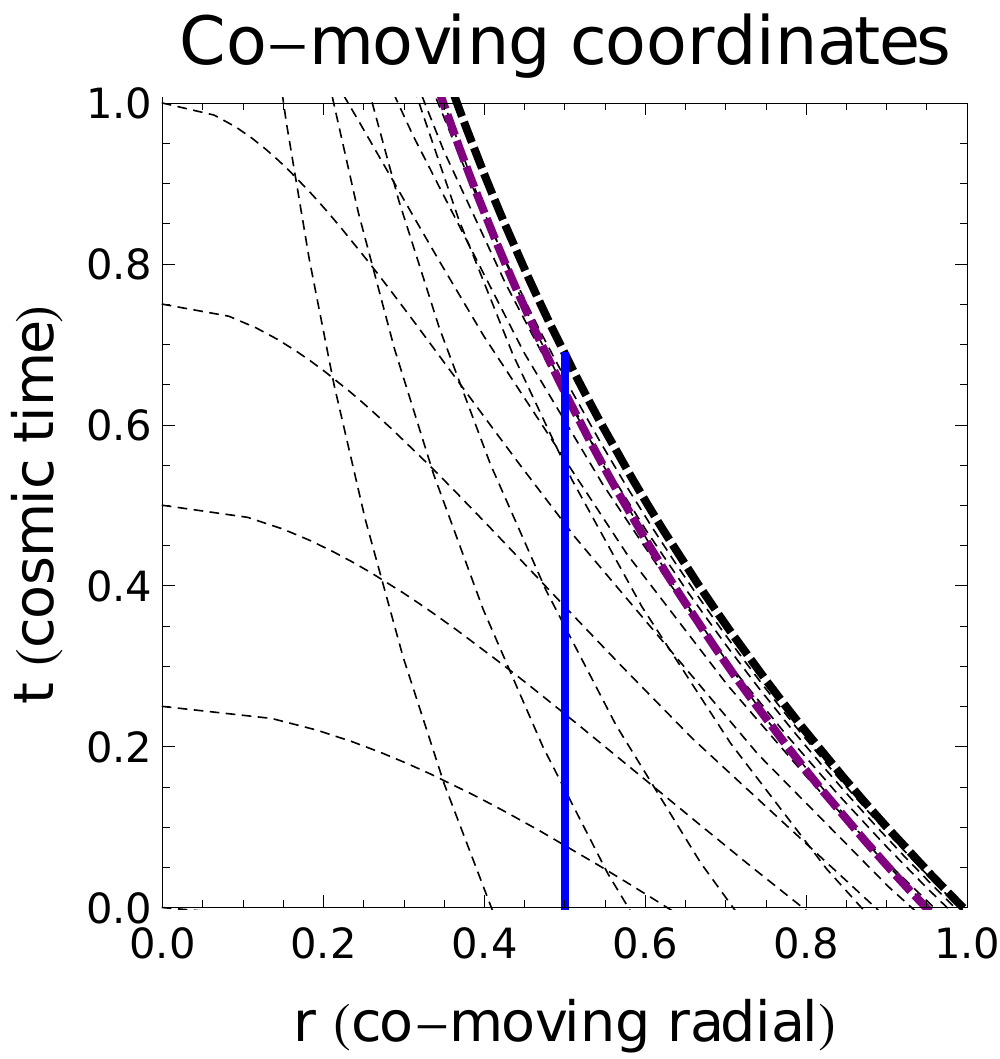}
&
\includegraphics[scale=0.7]{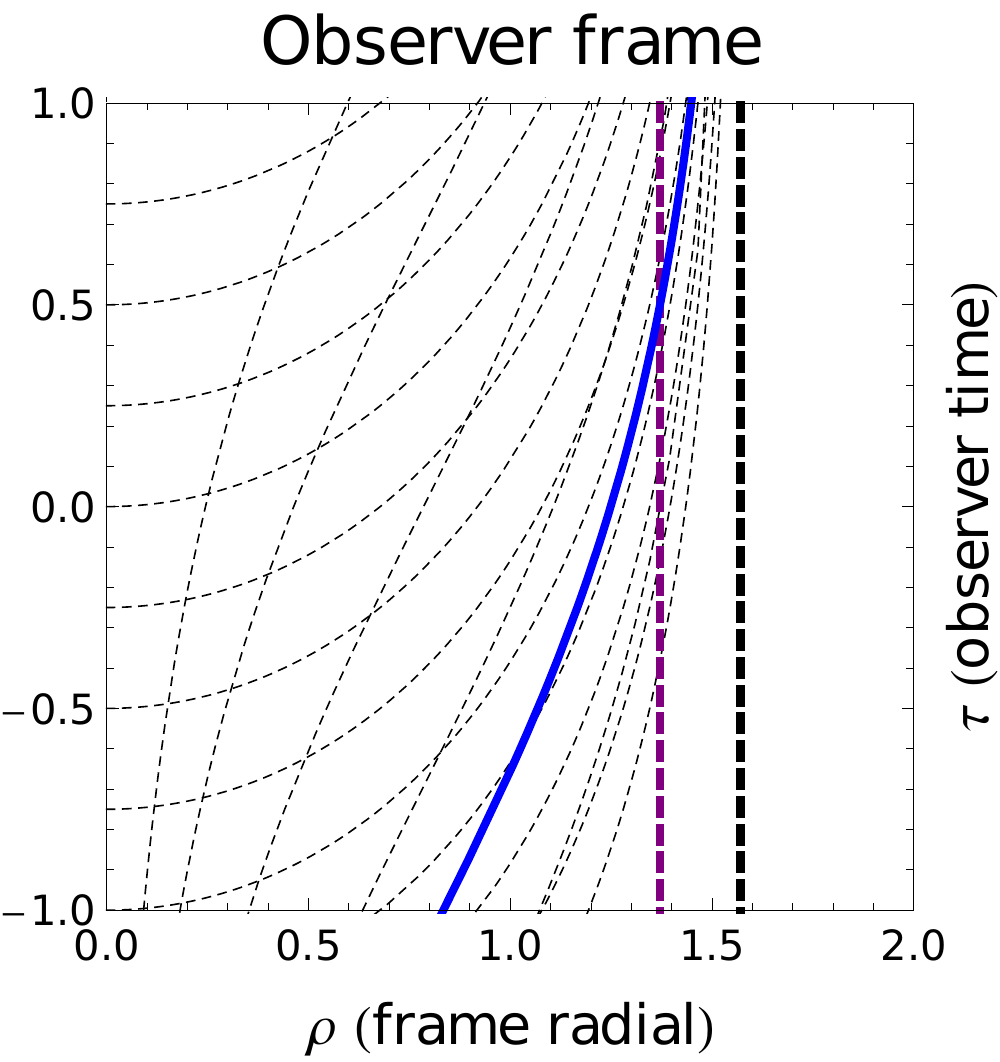} \\
\includegraphics[scale=0.7]{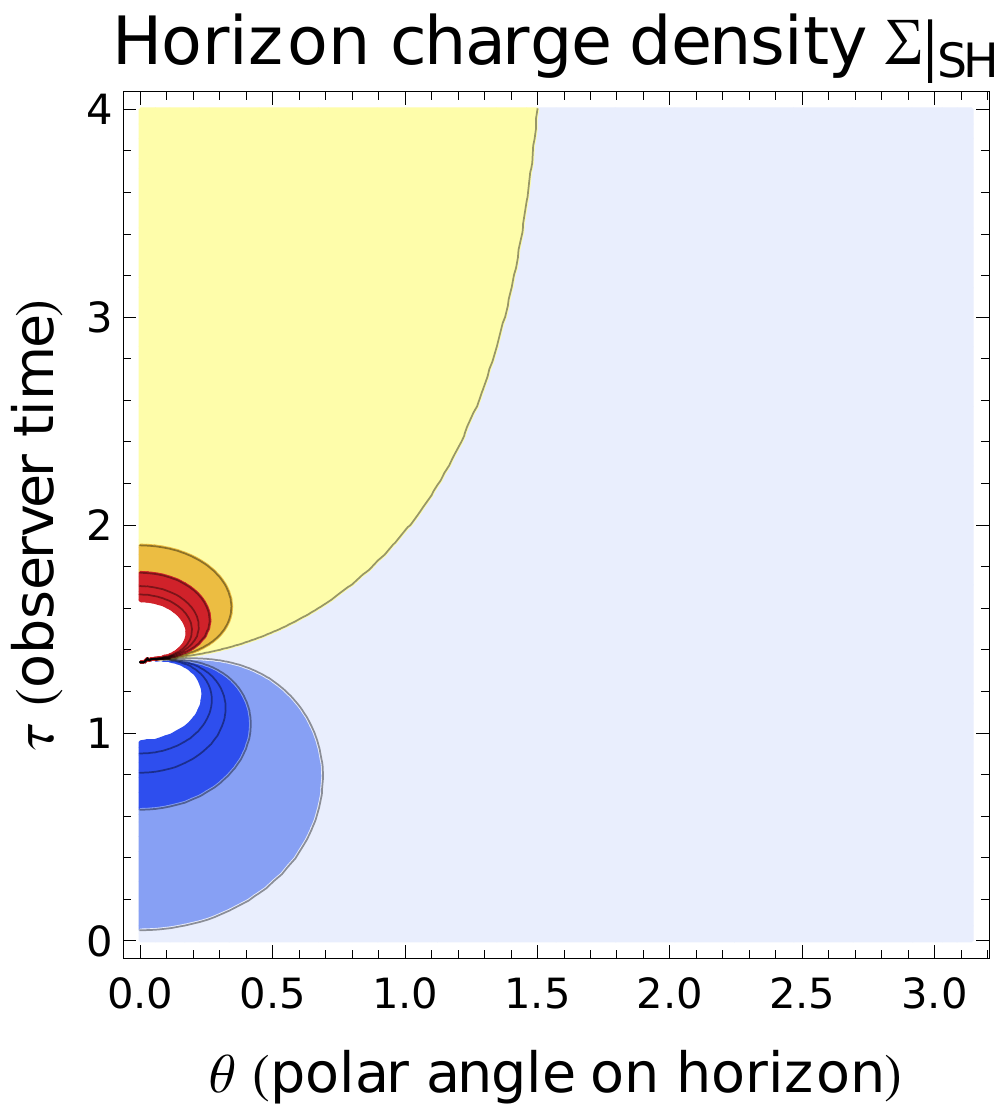}
&
\includegraphics[scale=0.725]{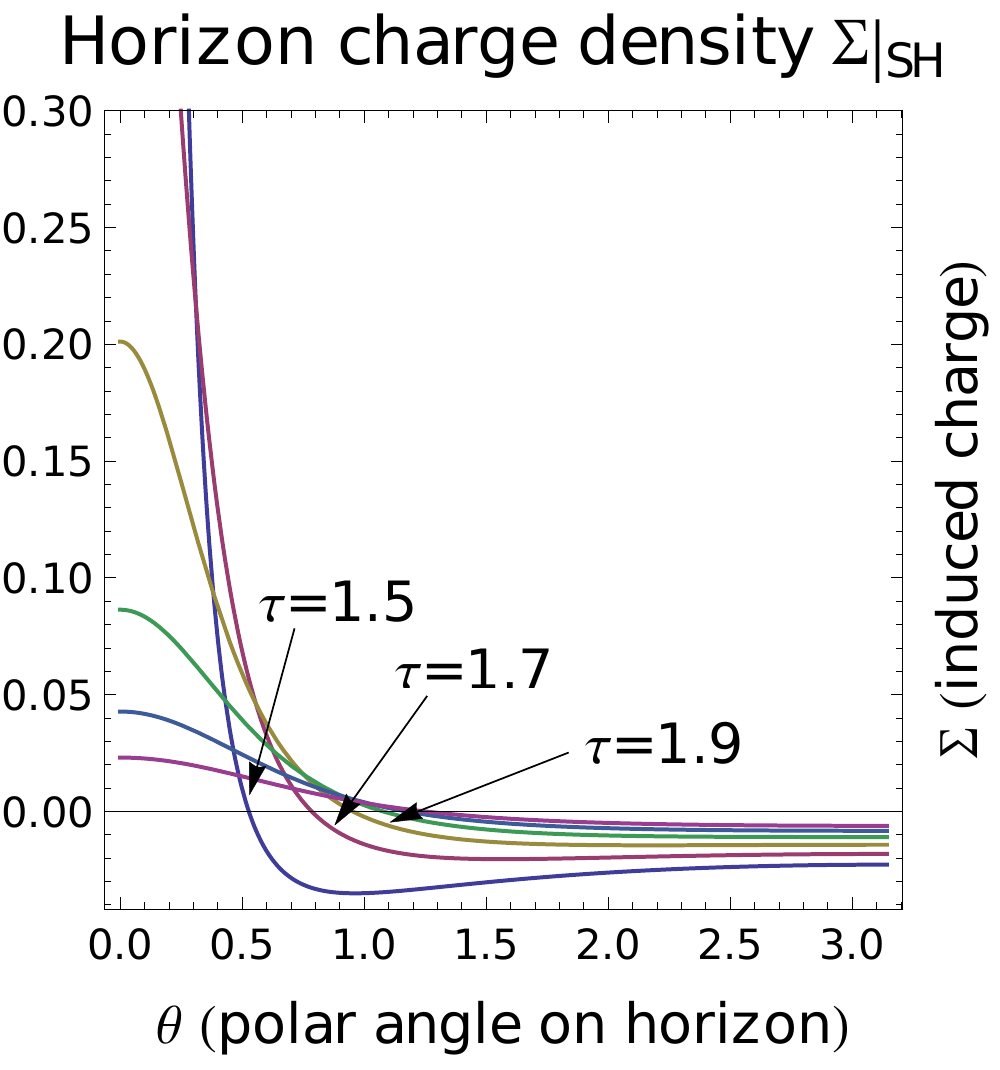}
\end{array}$
\end{center}
\caption{Top: Co-moving and frame coordinates in exponential inflation with $H_0 = 1, a_0=1, t_0=0$. Curves orthogonal to $\Oi$ (left vertical axis) are the spacelike slices of constant observer time $\tau$ (left) or cosmic $t$ (right), while the curves orthogonal to these denote constant radial distance. In this case the lines of constant frame radius are also at constant redshift parameter (\ref{dstransform}). The thick black and purple curves are $\Oi$'s event and stretched horizon, respectively, and the blue curve is the worldline of the charge $Q$. Bottom: Angular distribution of induced charge on the stretched horizon. In the left figure, blue means negative and yellow means positive induced charge.}
\label{inflationplots}
\end{figure}

A period of exponential inflation is described by the FRW metric (\ref{frwmetric}), with the scale factor and its inverse
\be
a(t) = a_0 e^{H_0 (t-t_0)}, \ \ b(a) = H_0^{-1} \ln a/a_0 + t_0.
\ee
It is convenient to leave $a_0$ and $t_0$ as free parameters so we can match to another cosmological epoch. From these formulas one can easily find explicit expressions for the frame coordinates. Using (\ref{tcoord}), (\ref{rhocoord}), and (\ref{rcoord}) we get
\be
\label{dstransform}
t(\tau,\sigma) = \tau - H_0^{-1} \ln \sqrt{\sigma}, \ \ r(\tau,\sigma) = \frac{\sqrt{\sigma-1}}{a(\tau) H_0}, \ \ \rho(\tau,\sigma) = H_0^{-1} \sec^{-1} \sqrt{\sigma}.
\ee
In this example one can easily invert the time-independent function $\rho = \rho(\sigma)$ to obtain $\sigma(\rho)$; plugging this into the formulas for $t,r$ then gives an explicit coordinate transform purely in terms of the frame coordinates $\tau,\rho$. Although $\rho$ can always be inverted like this in principle, it is hard to find examples where one can do it in terms of elementary functions. Using (\ref{fwmetric}), (\ref{gtautau}), and our result above for $r(\tau,\sigma)$ we have
\be
\begin{split}
ds^2 & = -\frac{d\tau^2}{\sigma} + d\rho^2 + \frac{\sigma-1}{H_0^2} d\Omega^2 \\
& = -\cos^2 (H_0 \rho) d\tau^2 + d\rho^2 + H_0^{-2} \sin^2 (H_0 \rho) d\Omega^2.
\end{split}
\ee
In writing the second line we used the inverse of $\rho$. We have obtained the static de Sitter metric, as one would expect.\cite{Klein:2010rk}

Clearly the event, particle and apparent horizons all occur at $\sigma \rightarrow \infty$ or $H_0 \rho = \pi/2$ as one expects. One can also obtain this result from (\ref{easyhorizon}). The proper area of all of these horizons is constant and given by $A_{horizon} \equiv 4\pi H_0^{-2}$. Indeed the proper area of any sphere at constant redshift $\sigma$ is constant in time, $A(\tau,\sigma) = 4\pi R^2(\tau,\sigma) \equiv 4\pi H_0^{-2} (\sigma-1)/\sigma$.

Now consider our falling charge again. We can read off the angular charge distribution on the horizon using (\ref{chargedensity}). Inserting (\ref{deltar}), (\ref{dstransform}), and replacing $r_Q$ with (\ref{rqframe}) we get the charge density:
\be
\label{chargedensity-ds}
\Sigma = -\frac{Q}{4\pi H_0^{-2}}  \frac{\sigma\left[ s - e^{H_0 \tau} s_Q \cos \theta \right]}{\left[ s^2 - 2 e^{H_0 \tau} s s_Q \cos \theta + e^{2 H_0 \tau} s_Q^2 \right]^{3/2}} \Bigg|_{horizon},
\ee
where we defined $s = \sqrt{\sigma-1}, s_Q  = \sqrt{\sigma_Q-1}$ and set $a_0=1, t_0=\tau_0=0$ for brevity. Placing the charge on the observer's worldline $\sigma_Q \rightarrow 1$ gives the correct static Coulomb field. We also have that for any $\sigma_Q$, the charge distribution on the true event horizon $\sigma \rightarrow \infty$ is just $\Sigma \equiv -Q/4\pi H_0^{-2}$ as explained above. 

The stretched horizon is a timelike surface very near the event horizon. In this formula this means we set $\sigma=\sigma_{SH} < \infty$ to some large but finite value. We can see what happens in fig. \ref{inflationplots}. While the charge is in view it induces a negative charge $Q_{ind} = -Q$ across the horizon. As it passes through the horizon it induces a large spike of positive charge which then spreads exponentially fast across the top half of the horizon, leaving an overall neutral, symmetric dipole after about a scrambling time of order $H_0^{-1}$. For example a charge today would take on the order of $10^{10}$ years to spread across an order one fraction of the horizon while during primordial inflation it would have taken no longer than about $10^{-25}$ seconds (for $H_{inf} \sim$ 1 GeV).

To connect explicitly to known results in the literature, we note that in the Rindler limit, one simply sees the induced charge spread out exponentially for all time, because the horizon is a plane. The picture here is refined by the constraint of the Gauss law: after the charge passes outside the horizon it must induce a net charge of zero. This is consistent with our identification of the charge density with the bulk data of the point charge $Q$: no matter how we count things the total charge of the system is always $Q_{total} = Q_{bulk} + Q_{horizon} = 0$.

\subsection{Power law scale factors ($-1 < w \leq 1$)}
\label{powerlawsection}

\begin{figure}[h]
\begin{center}$
\begin{array}{cc}
\includegraphics[scale=0.7]{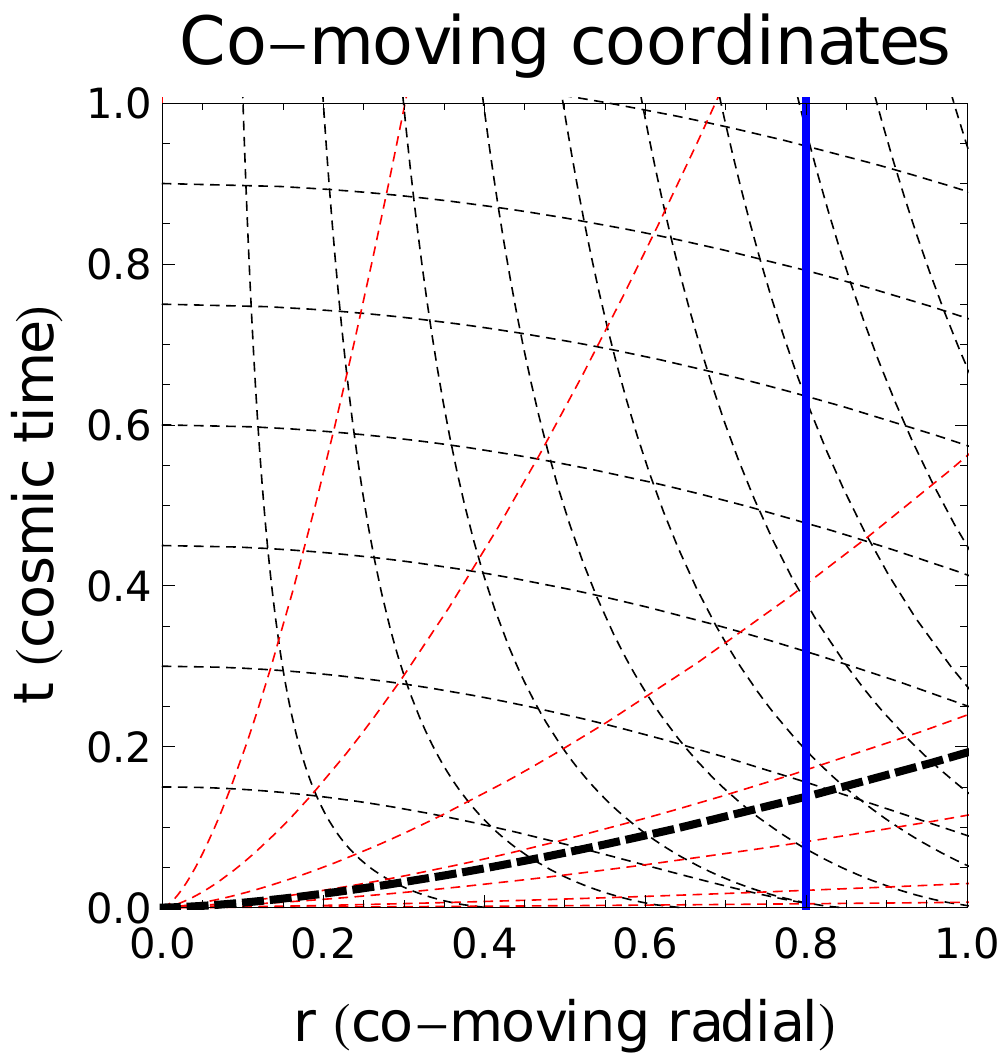}
&
\includegraphics[scale=0.68]{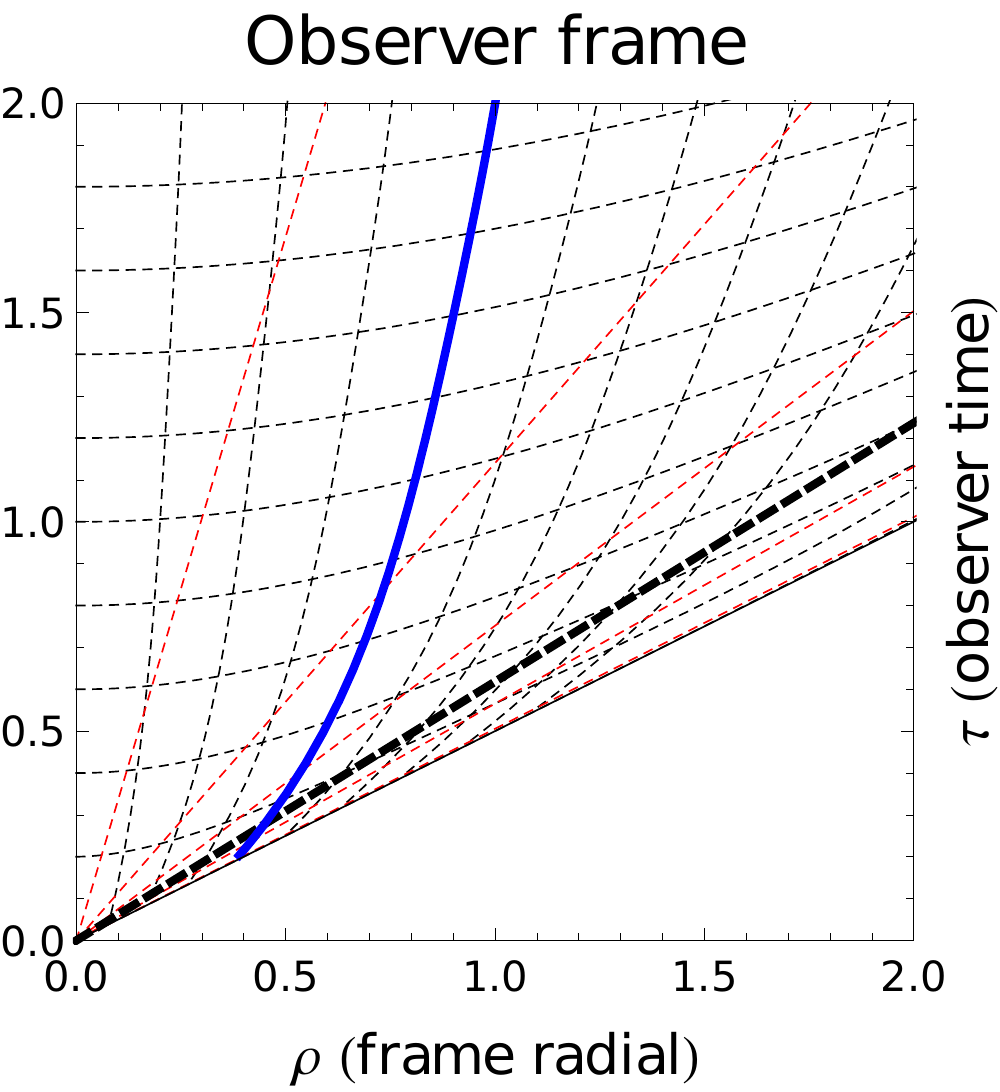}\\
\includegraphics[scale=0.7]{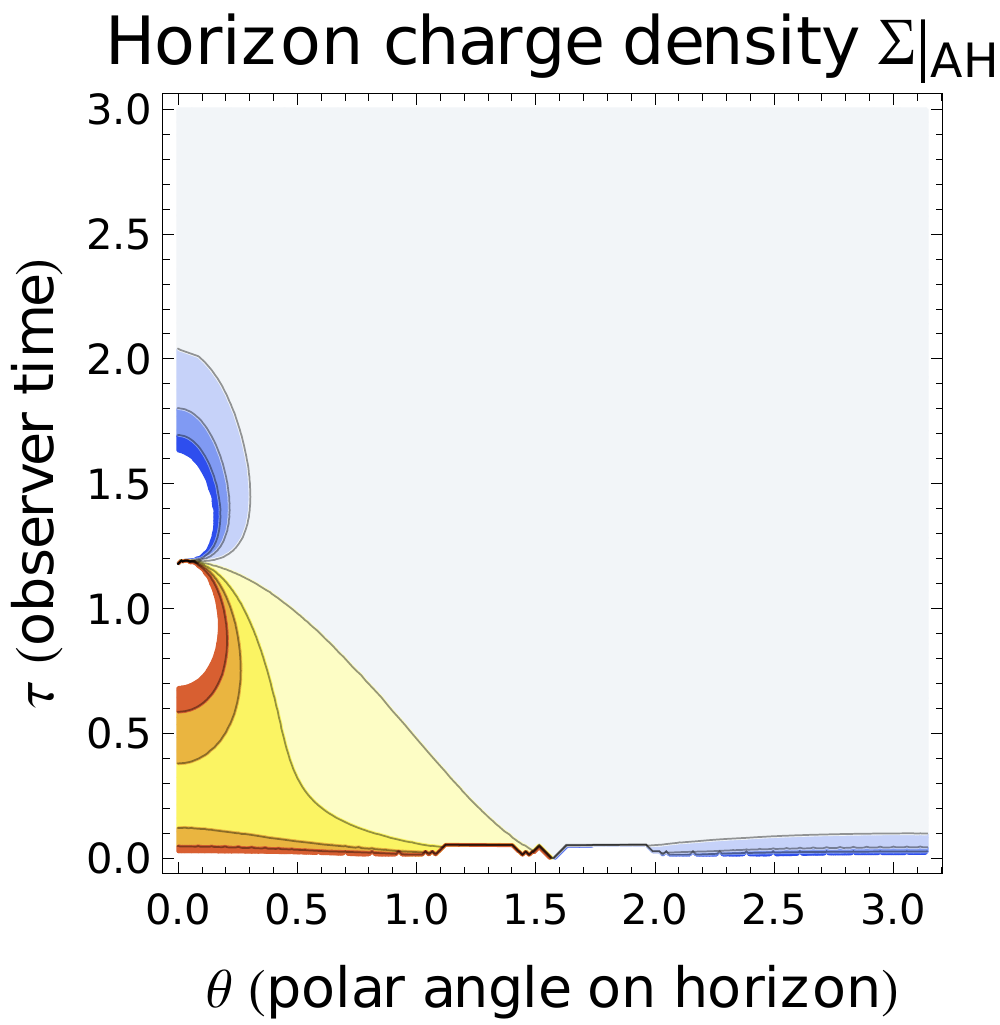}
&
\includegraphics[scale=0.7]{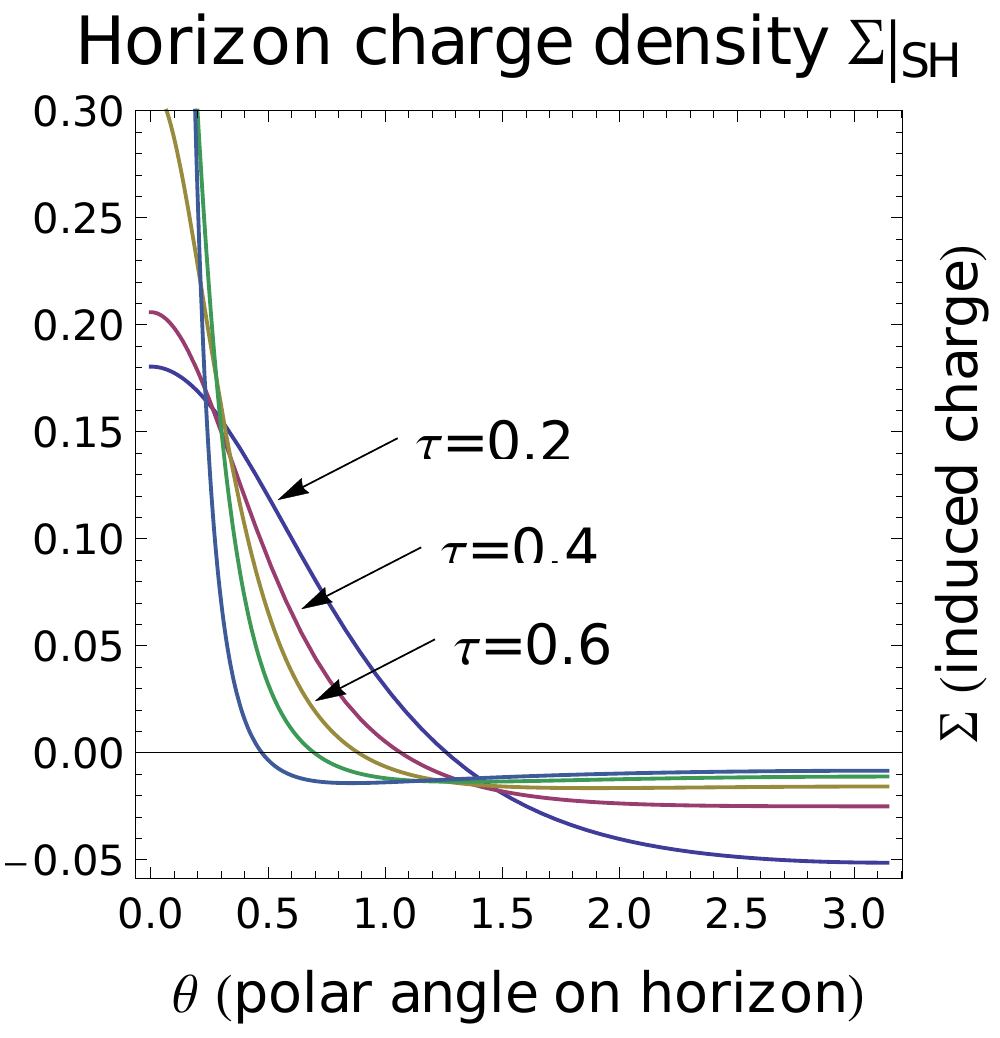}
\end{array}$
\end{center}
\caption{Top: Co-moving and frame coordinates in a kinetic-energy dominated big bang cosmology with $a_0=1, t_0=1$, in the same notation as fig. \ref{inflationplots}. In red we have also plotted some lines of constant redshift parameter ($\sigma=1.01,1.05,1.2,1.5,2,4,10$). The thick black curve is the apparent horizon. Bottom: Induced charge distribution on the apparent horizon, with $r_Q = 10$.}
\label{kineticchargeplots}
\end{figure}

Another set of simple and relevant examples are the big bang cosmologies with power-law scale factors $a \sim t^{\alpha}$. A generic value of $\alpha$ gives transformation laws in terms of some hypergeometric functions, but much of the physics is transparent. Because we want to understand scrambling in decelerating cosmologies, we take a kinetic-energy dominated universe $\alpha = 1/3$ for concreteness, but the generalization is obvious in principle. The scale factor is
\be
a(t) = a_0 \left(\frac{t}{t_0}\right)^{1/3}, \ \ b(a) = t_0 \left( \frac{a}{a_0} \right)^{3}.
\ee
Here again $a_0, t_0$ are free parameters. Using (\ref{tcoord}), (\ref{rhocoord}), and (\ref{rcoord}), we get the usual transformations
\be
\label{kineticcoords}
t(\tau,\sigma) = \tau \sigma^{-3/2}, \ \
r(\tau,\sigma) = \frac{3 \tau}{a(\tau)} \sqrt{\frac{\sigma-1}{\sigma}}, \ \
\rho(\tau,\sigma) = \tau \frac{1+2\sigma}{\sigma} \sqrt{\frac{\sigma-1}{\sigma}}.
\ee
Again using (\ref{fwmetric}), (\ref{gtautau}) we find that the metric takes the simple form
\be
ds^2 = -\left(\frac{2\sigma-1}{\sigma}\right)^2 d\tau^2 + d\rho^2 + \left( \frac{3 \tau}{\sigma} \right)^2 (\sigma-1) d\Omega^2.
\ee

Contrary to the de Sitter case, $g_{\tau\tau}$ is finite as $\sigma \rightarrow \infty$, reflecting the fact that this spacetime has no event horizon. In the same limit, the spatial spheres shrink to zero radius: this is the big bang. The radius of the spatial sphere at any fixed redshift grows linearly in observer time. In particular, using (\ref{easyhorizon}) one finds easily that the apparent horizon is located at constant redshift parameter $\sigma_{AH} = (1+ \sqrt{5})/2$.

Consider once again our free-falling charge. We can read off the charge density with (\ref{chargedensity}). Replacing $r_Q$ with (\ref{rqframe}) we get:
\be
\label{chargedensity-kinetic}
\Sigma = -\frac{Q}{4\pi} \frac{\sigma}{(3\tau)^2} \frac{ s - (\tau/\tau_0)^{-2/3} s_Q \cos \theta }{\left[ s^2 - 2 (\tau/\tau_0)^{-2/3} s s_Q \cos \theta + (\tau/\tau_0)^{-4/3} s_Q^2 \right]^{3/2}} \Bigg|_{horizon},
\ee
where this time we have defined $s=\sqrt{(\sigma-1)/\sigma}$. We can easily check some simple limits again. Placing the charge on the observer's worldline $\sigma_Q \rightarrow 1$, we find a Coulomb field redshifting in time,
\be
F_{\tau\rho} \bigg|_{\sigma_Q \rightarrow 1} = -\frac{Q}{4\pi} \frac{1}{(3\tau)^2}.
\ee
Meanwhile the spatial spheres have area growing at precisely the right rate to cancel this effect, so that we still satisfy the Gauss law. Since this spacetime has no event horizon, the boundary conditions on the field are simply that $F_{\tau\rho} \rightarrow \infty$ at the big bang $\sigma \rightarrow \infty$, which is certainly satisfied.\footnote{One can see that this is the right boundary condition by again appealing to the Gauss law.}

From these formulas and fig. \ref{kineticchargeplots} it is clear what is going on. Consider a configuration in which the image on the horizon is already scrambled into a neutral dipole. By the method of images this is obviously equivalent to a point charge $Q$ starting behind the horizon. The observer $\Oi$ will see his horizon grow and the charge fall away, but the horizon grows faster. Thus he sees the charge's image on the horizon coalesce or ``de-scrambles'' from a dipole back into a point charge which then re-appears inside the horizon. This occurs at a power-law rate as one can see easily from (\ref{chargedensity-kinetic}).

%We can write everything explicitly in terms of $\rho$ by inversion. What we want to do is invert $\rho$ for fixed $\tau$ to get $\sigma(\tau,\rho)$. This amounts to solving a cubic equation and can be done explicitly. We want to solve
%\be
%\sigma^3 \rho^2/\tau^2 = (1+2\sigma)^2 (\sigma-1)
%\ee
%for $\sigma$. This cubic has positive discriminant (see wiki) for $\rho/\tau > 0$ so it has three real roots. However at $\rho=0$ one can find easily that the roots are $-1/2,-1/2,$ and $1$; the third branch is the one we want (i.e. we demand $\sigma(\tau,\rho=0) = 1$). With a bit of work one arrives at the following answer: let
%\be
%\label{kineticbeta}
%B(\tau,\rho) = \frac{2}{\sqrt{4-\rho^2/\tau^2}}, \ \ \beta(\tau,\rho) = \frac{1}{3} \tan^{-1} \left( 2 \sqrt{B^2-1} \right),
%\ee
%then the correct branch for $\sigma$ is
%\be
%\label{kineticsigma}
%\sigma(\tau,\rho) = B \cos \beta.
%\ee
%This function has the right physical properties: for any fixed $\tau$ we see that $\sigma$ runs smoothly from $1$ to $\infty$ as $\rho$ runs from $0$ to $2\tau$; the big bang is a distance $\rho_{BB}(\tau) = 2 \tau$ from the observer at time $\tau$.\footnote{This matches Klein-Randles eq. (77).-- I don't like the BB/event horizon dichotomy; I think the general statement is that the past boundary of the past lightcone of $\Oi$ is at $\sigma \rightarrow \infty$...} To check that (\ref{kineticsigma}) really is the inverse of $\rho$ given by (\ref{kineticcoords}), a very useful identity is that for any $x$,
%\be
%4 \cos^3 \left( \frac{x}{3} \right) - 3 \cos \left( \frac{x}{3} \right) = \cos \left( x \right).
%\ee

\subsection{Junctions of epochs}

\begin{figure}[h]
\begin{center}$
\begin{array}{cc}
\includegraphics[scale=0.72]{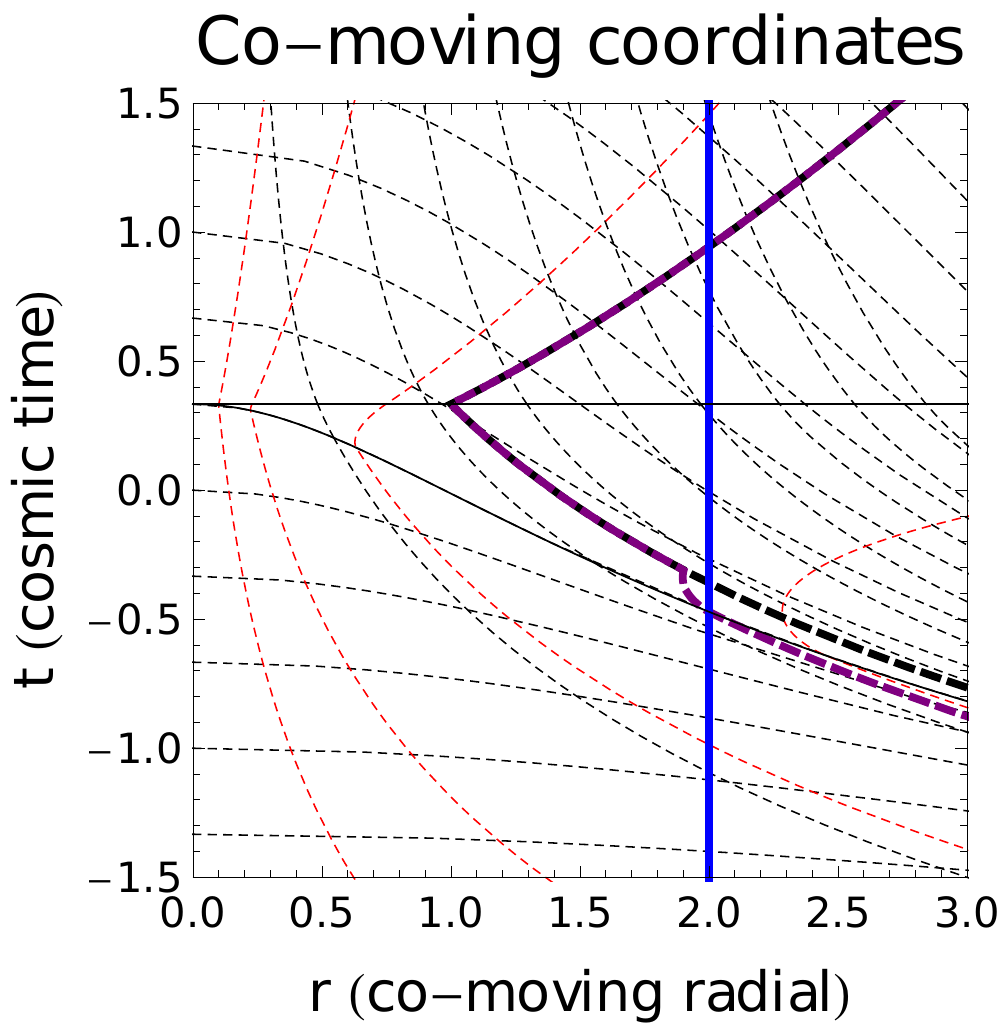}
&
\includegraphics[scale=0.66]{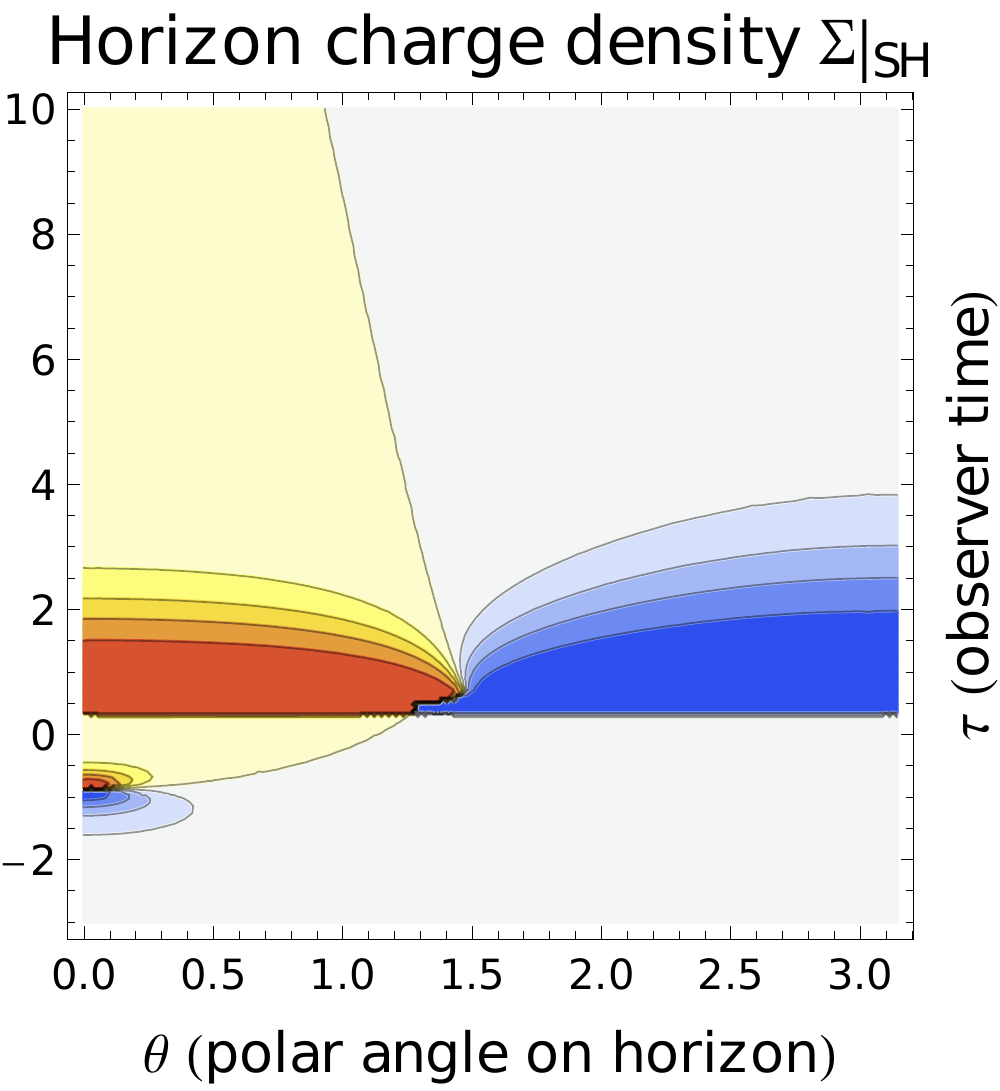}
\end{array}$
\end{center}
\caption{Left: co-moving and frame coordinates in a universe that inflates with Hubble constant $H_0=1$ followed by a kinetic dominated phase $\alpha=1/3$ after $t=t_0=\alpha H_0^{-1}$. As indicated in the text, the coordinates are broken into three regions by the solid black lines. Right: Induced charge density on the stretched horizon, with $r_Q$ tuned so that the charge scrambles within about an e-folding of the end of inflation. Here we are plotting $\arctan{\Sigma}$ for graphical clarity: the stretched horizon moves inward very rapidly at $t=t_0$ and this causes a large spike in the induced charge.}
\label{junctionplots}
\end{figure}

\begin{figure}[h]
\begin{center}$
\includegraphics[scale=1.1]{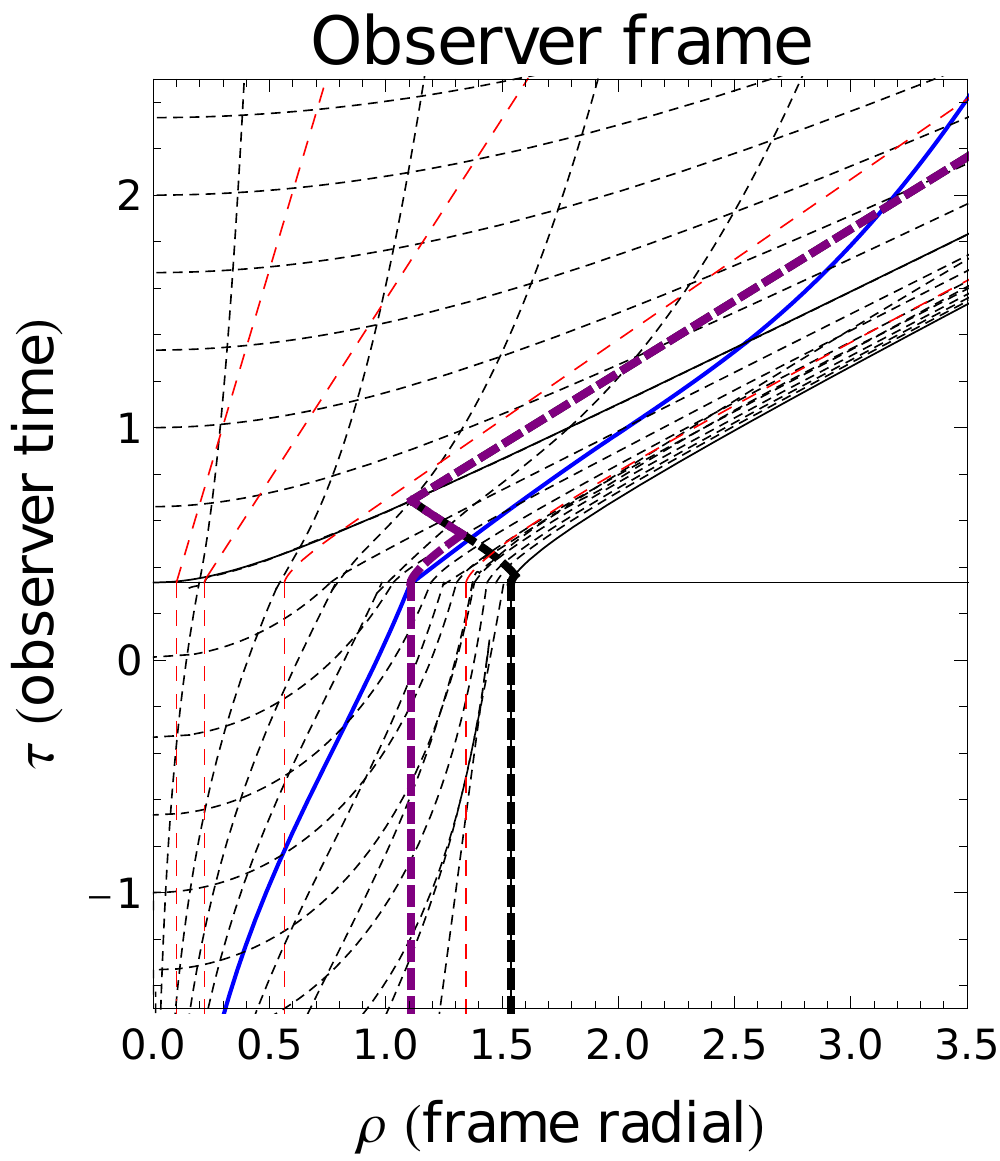}$
\end{center}
\caption{Observer frame in our junction cosmology. The thin black lines are lines with constant FRW $t$ or $r$ so in particular the vertical ones are co-moving geodesics. Thin red lines are lines of constant redshift parameter $\sigma=1.01,1.05,1.4,20$. The blue line is again some representative charge worldline. We have severely stretched the horizon in purple for graphical clarity. The coordinates during the inflationary period are bounded by the apparent horizon and then ``grow'' during the decelerating phase, and we can clearly see the charge spend some transient period behind the horizon. The detailed behavior near $\tau=\tau_0$ smooths over some numerical error.}
\label{junctionframeplot}
\end{figure}

When we measure cosmological perturbations from the early inflationary era, we view them after they pass through some of the later cosmological evolution. We are thus interested in studying the precise evolution of the perturbations, or more generally any observables, through some combination of cosmological epochs.

Continuing in the vein of the previous two sections, we focus on a universe which is exponentially inflating at early times and then exits into a kinetic energy-dominated phase. The latter has a decelerating scale factor, $a(t) \sim t^{1/3}$. We will see that a charge ``scrambled'' onto the horizon during the accelerating phase is later ``de-scrambled'' as it re-enters the horizon during the decelerating phase.

We can formulate the problem in some generality. Consider a universe which we divide into two periods around some time $t_0$. The scale factor is
\be
a(t) = \begin{cases} a_E(t), \ \ t \leq t_0 \\ a_L(t), \ \ t \geq t_0. \end{cases}
\ee
for example a period of inflation followed by some power law
\be
a_E(t) = a_0 \exp H_0 (t-t_0), \ \ a_L(t)=a_f (t/t_0)^{\alpha}.
\ee
Although $a(t)$ should in reality be smooth, it is convenient to allow for a junction where some derivatives are discontinuous. In our example we can satisfy continuity of the zeroth and first derivatives by $a_f = a_0, \ \ t_0 = \alpha H_0^{-1}$. If we want a decelerating phase $\alpha < 1$ then the second derivative is necessarily discontinuous (since the universe abruptly switches from accelerating to decelerating). This can be accounted for by a thin shell of stress-energy at the junction, by the Israel matching conditions.

Some care has to be taken in working out the geometry and electrodynamics. In particular we have to be careful when inverting the scale factor to get $b=b(a)$. This function will also have discontinuous second derivative, but all the coordinate transforms are perfectly continuous. Since $a(t)$ is monotonic we have $a \leq a_0$ for $t \leq t_0$ and the same with the inequalities reversed. In our example
\be
b(a) = \begin{cases} b_E(a) = H_0^{-1} \ln (a/a_0) + t_0, &a \leq a_0 \\
b_L(a) = H_0^{-1} (a/a_0)^{3}/3, &  a \geq a_0. \end{cases}
\ee
Clearly $b$ and $b'$ are continuous at $a=a_0$, but not $b''$.

Let us work out the transformations (\ref{tcoord}), (\ref{rcoord}). Note that $t$ and $\tau$ coincide on $\Oi$'s worldline, so we write $t_0 = \tau_0 = H_0^{-1}/3$. Composing $b$ with $a(\tau)/\sqrt{\sigma}$ breaks up the $\tau-\sigma$ plane into three regions (early, middle and late) as shown in figure \ref{junctionplots}. For any early frame time $\tau \leq \tau_0$ and any $1 \leq \sigma \leq \infty$ it is clear that we can put $b(a(\tau)/\sqrt{\sigma}) = b_E(a_E(\tau)/\sqrt{\sigma})$. In the late region $\tau \geq  \tau_0 \sigma^{3/2}$ we need $b_L$ and $a_L$. The middle region $\tau_0 \leq \tau \leq \tau_0 \sigma^{3/2}$ is the subtle one: we need to use $b_E$ but $a_L$ because $\sigma \geq 1$. In all we find the coordinate transformation for FRW time $t$ by
\be
t(\tau,\sigma) = \begin{cases} 
\tau - H_0^{-1} \ln \sqrt{\sigma} & \tau \leq \tau_0 \\
H_0^{-1} \left[ (1+ \ln(\tau/\tau_0))/3-\ln \sqrt{\sigma}\right], & \tau_0 \leq \tau \leq \tau_0 \sigma^{3/2} \\
\tau \sigma^{-3/2}, & \tau \geq  \tau_0 \sigma^{3/2}.
\end{cases}
\ee
It is instructive to check that this function is continuous in both variables. To get the co-moving radial coordinate $r=r(\tau,\sigma)$ we have to do some integrals along the spacelike geodesics, breaking up the domains in the $\tau-\sigma$ plane in the same way. One finds that
\be
r(\tau,\sigma) = \frac{1}{a(\tau) H_0} \begin{cases}
\sqrt{\sigma-1}, & E \\
\frac{\tau}{\tau_0} \sqrt{\frac{\sigma_*-1}{\sigma_*}} + \sqrt{\sigma-1} - \sqrt{\sigma_*-1}  & M \\
\frac{\tau}{\tau_0} \sqrt{\frac{\sigma-1}{\sigma}}, & L
\end{cases}
\ee
where $\sigma_* = \sigma_*(\tau)$ is the redshift parameter at which the spatial geodesic orthogonal to $\Oi(\tau)$ crosses the junction,
\be
\frac{\tau}{\tau_0} = \sigma_*^{3/2}.
\ee
Finally, using (\ref{gtautau}) and (\ref{gsphere}) we can write down the metric coefficients (\ref{fwmetric}) in the frame. Being careful with the domains, one finds that $g_{\tau\tau} = -1/\sigma$ in the early region, $-((2\sigma-1)/\sigma)^2$ in the late region, and
\be
g_{\tau\tau} = \left[ \left( \frac{\tau}{\tau_0} \right)^{-1} \sqrt{\sigma} + 2\sqrt{\frac{\sigma-1}{\sigma}} \sqrt{\frac{\sigma_*-1}{\sigma_*}} - \left( \frac{\tau}{\tau_0} \right)^{-1} \sqrt{\frac{\sigma-1}{\sigma}} \left(\sqrt{\sigma-1} -\sqrt{\sigma_*-1} \right) \right]^2
\ee
in the middle region. Once again, the early and late regions match the results from the previous sections, and $g_{\tau\tau}$ is continuous everywhere.

Since the scale factor is not accelerating in the future, our observer $\Oi$ does not have an event horizon. However, he does see an apparent horizon. At early times he might have mistaken it for the de Sitter horizon, but after the phase transition he will see the horizon recede and grow in area, with $R_{AH} \rightarrow \infty$ as $t,\tau \rightarrow \infty$. To be precise we can solve (\ref{easyhorizon}) to find the redshift parameter of the apparent horizon: at early times we have $\sigma_{AH} = \infty$, and in the late region $\sigma_{AH} = (1+\sqrt{5})/2$, in accordance with our earlier results. During the middle period we have
\be
\sigma_{AH}(\tau) = \frac{1}{4} \left( \frac{ A^2(\tau) + 1}{A(\tau)} \right)^2, \ \ A = \frac{\tau}{\tau_0} \sqrt{\frac{\sigma_*-1}{\sigma_*}} - \sqrt{\sigma_*-1}.
\ee
It is easy to check that this continuously interpolates between the early and late periods.

As usual we can read off the charge density from (\ref{chargedensity}). To keep the expressions tractable, put $s_E = \sqrt{\sigma-1}$ and $s_L = \sqrt{(\sigma-1)/\sigma}$ as before, and use (\ref{rqframe}) to write $r_Q = s_Q/a_0 H_0$ with $s_Q = \sqrt{\sigma_Q-1}$. Then at early times we have
\be
\Sigma_E = -\frac{Q \sigma}{4\pi H_0^{-2}} \frac{s_E - e^{H_0 (\tau-\tau_0)} s_Q \cos \theta}{\left[ s_E^2 - 2 e^{H_0 (\tau-\tau_0)} s_E s_Q \cos \theta + e^{2 H_0 (\tau-\tau_0)} s_Q^2 \right]^{3/2}},
\ee
and at late times
\be
\Sigma_L = -\frac{Q \sigma}{4\pi H_0^{-2} \left( \frac{\tau}{\tau_0} \right)^2} \frac{s_L - (\tau/\tau_0)^{-2/3} s_Q \cos \theta }{\left[ s_L^2 - 2 (\tau/\tau_0)^{-1/3} s_L s_Q \cos \theta + (\tau/\tau_0)^{-2/3} s_Q^2 \right]^{3/2}},
\ee
in agreement with (\ref{chargedensity-ds}) and (\ref{chargedensity-kinetic}), respectively, where again we set $a_0=1$. During the middle period we have the somewhat more complex behavior
\be
\Sigma_M = -\frac{Q \sigma}{4\pi H_0^{-2} \left( \frac{\tau}{\tau_0} \right)^2} \frac{s_M-(\tau/\tau_0)^{-2/3} s_Q \cos \theta}{ \left[ s_M^2 - 2 (\tau/\tau_0)^{-1/3} s_M s_Q \cos \theta + (\tau/\tau_0)^{-2/3} s_Q^2 \right]^{3/2}},
\ee
where
\be
s_M = s_M(\tau,\sigma) = \sqrt{\frac{\sigma_*-1}{\sigma_*}} + \left( \frac{\tau}{\tau_0} \right)^{-1} \left[ \sqrt{\sigma-1} - \sqrt{\sigma_*-1} \right].
\ee
Here, the charge densities are evaluated on the horizon. In the middle and late regions $\tau>\tau_0$ the apparent horizon is spacelike and we can set $\sigma = \sigma_{AH}(\tau)$ directly. At early times $\tau \leq \tau_0$ the horizon is null, so we need to stretch it by placing it at some large finite redshift $\sigma_{SH} <\infty$. Once again $\Sigma = \Sigma(\tau,\sigma)$ is continuous and so is the redshift of the apparent horizon, so we have a continuously varying image on the horizon through the entire cosmic history.

Following the earlier sections, the interpretation is clear. A charge $Q$ even slightly displaced from the observer $\Oi$ which begins inside the horizon during the early period of inflation will, if inflation lasts long enough, fast-scramble onto the apparent horizon. However, in the later decelerating period, the image will then de-scramble at a power law rate as the point charge reappears inside the horizon.

\section{Conclusions}
\label{conclusionsection}
In this paper, we have studied the scrambling and de-scrambling of point charges across the apparent horizon in a flat FRW universe. We showed how to do this in terms of a reference frame based on an inertial observer, referring only to events to which he has causal access. We found that accelerating cosmologies scramble the charge as it falls past their horizons, while decelerating cosmologies de-scramble the previously-scrambled images of charges back into a local bulk charge at ``horizon re-entry''. We showed that these processes occur at slower-than-exponential rates except during de Sitter-like epochs.

Following \cite{Sekino:2008he,Susskind:2011ap}, this suggests that it may be possible to find a holographic description of observations made in cosmology in terms of locally interacting degrees of freedom, except possibly during exponential inflation. The calculations presented here should be of significant value in the search for an explicit holographic model of cosmological spacetimes.

Our approach is very robust at the level it has been formulated. To keep the discussions concrete, we have only worked with matter consisting of a classical Maxwell field with point charge source. However, the arguments are very basic, relying almost entirely on causal structure. In particular, we have at no point used the Einstein equations (other than to motivate the use of various scale factors): our results hold in any theory admitting FRW solutions, like the one in which we live. Similar conclusions should hold for neutral massive particles,\cite{Thorne:1986iy} although in that case one studies the propagation of stress-energy on the stretched horizon.

Clearly the next step is to study quantum mechanical effects. An obvious candidate is to understand the spectrum of inflationary perturbations. In particular, it would be very interesting to get a sharp understanding of various ``conservation outside the horizon'' laws (eg. \cite{Weinberg:2003sw}) from this viewpoint. We leave this to future work.

Ultimately, any complete quantum theory of gravity should have a low-energy limit which describes our observed universe to high accuracy. This theory should be unitary, at least to very high precision at low energies. The physically relevant formulation of unitarity is that the probabilities of all measurement outcomes sum to one. In this paper we have shown how to systematically formulate classical physics on precisely the set of spacetime events that a physical observer can probe. This framework thus gives a very natural way to implement unitarity. We have obtained a very concrete suggestion about low-energy bulk information near horizons: it localizes or de-localizes in time in order to remain accessible to the observer.

\acknowledgments{
We thank Mio Alter, Tom Banks, Brandon diNunno, Sandipan Kundu and especially David Klein and Tom Mainiero for helpful conversations, and Elizabeth Maloy for help with illustrations. This material is based upon work supported by the National Science Foundation under grant numbers PHY-1316033 and PHY-0969020.}

\appendix

\section{Causal structure of FRW}
\label{causalstructureappendix}

In a cosmological spacetime described by an FRW metric, arbitrary observers, including inertial ones, may experience horizons. In general they can observe particle, event, and apparent horizons. These horizons are generically distinct, with all three definitions coinciding precisely for an eternal de Sitter universe. 

The particle and event horizons are global structures which depend on the entire past or entire future of the spacetime, respectively. Events behind the particle horizon are those to which the observer cannot have sent a signal; events behind the event horizon are those from which he will never receive a signal. The intersection of the interior regions bounded by these horizons is the causal diamond of the obsever, the set of spacetime events which he can measure by probing and receiving a signal. We show the general situation in fig. \ref{causalstructurepicture}.

Consider an inertial observer $\Oi$ who, without loss of generality, can be taken to sit at $r=0$. The metric in co-moving coordinates is (\ref{frwmetric}), where we take flat spatial sections for simplicity. In terms of these coordinates, one can integrate the condition $ds^2=0$ for radial null rays to find that the event and particle horizons seen by $\Oi$ at any time $t$ are just the 2-spheres with co-moving radius
\be
\label{eventhorizonco-moving}
r_{EH}(t) = \int_{t}^{t_f} \frac{dt'}{a(t')}, \ \ r_{PH}(t) = \int_{t_0}^{t} \frac{dt'}{a(t')},
\ee
respectively. Here $t_0$ and $t_f$ are the co-moving time coordinates at the beginning and end of time, respectively; in a big bang cosmology we have finite $t_0$ (usually normalized to $t_0 = 0$) and $t_f \rightarrow \infty$, while in inflationary spacetimes one sometimes takes $t_0 \rightarrow -\infty$. These horizons exist if and only if these integrals converge; in particular there is an event horizon if and only if the scale factor is accelerating as $t \rightarrow \infty$, i.e. is growing at least as fast as $a(t) \sim t^{\alpha}$ with $\alpha > 1$. The proper radius of the event horizon $r^{prop}_{H}(t) = a(t) r_{EH}(t)$ is non-decreasing in time, and more generally obey the analogues of the laws of black hole thermodynamics.\cite{davies}

For any fixed angles, the curves $(t,r_{H}(t),\theta_0,\phi_0)$ are null geodesics, and letting the angles vary over the sphere, the codimension-1 hypersurface generated by these curves is a null hypersurface. Because these horizons form null hypersurfaces, $\Oi$ cannot observe anything dynamical happening on them, and so one can introduce a timelike hypersurface placed very near the physical horizons; this surface is called the stretched horizon.\cite{Thorne:1986iy,Susskind:1993if}

In addition to the particle and event horizons, $\Oi$ may also observe an apparent horizon. These are local (in time) and much less rigid structures than the global horizons explained above. For our purposes the apparent horizon at any time $t$ is the unique 2-sphere such that at least one of its four familes of surface-orthogonal null geodesics have vanishing geodesic expansion $\Theta$ (see eg. \cite{poissonbook}). Pick any spatial 2-sphere $S = S^2(t,r)$. We would like to compute the expansions $\theta$ of the four families of null geodesic congruences orthogonal to $S$. The tangent vector fields of these congruences can be written $k^{\mu} = (\pm_t a(t), \pm_r 1, 0, 0)$ where the two choices of sign label the four congruences. This is a non-affine parametrization, $\nabla_k k = \kappa k$ with $\kappa = \pm_t 2 \dot{a}(t)$. Using this we can write down the geodesic expansion of these families \cite{poissonbook}:
\be
\frac{\Theta}{2} = \pm_r \frac{1}{r} \pm_t \dot{a}.
\ee
From here it is easy to see that at time $t$ the spheres with $\theta=0$ will be at co-moving radial coordinate $r_{AH}(t) = - \pm_t \pm_r 1/\dot{a}(t)$. Obviously we need $r_{AH} > 0$, so we see that there will only be one apparent horizon
\be
r_{AH}(t) = \frac{1}{|\dot{a}(t)|} = \frac{1}{|a(t) H(t)|}
\ee
and in fact two of its null congruences will have $\Theta=0$. The case $\dot{a} = 0$ is just flat Minkowski space with the spatial coordinates scaled by $a$ so there is no apparent horizon (of course $\theta \rightarrow 0$ for some families as one sends $r \rightarrow \infty$ but there is no trapped region). 

The apparent horizon is again the hypersurface generated by the curves $(t, r_{AH}(t), \theta_0, \phi_0)$ for all angles. Crucially, this hypersurface \emph{need not be null}, that is, its normal can in general take any spacetime signature. In particular we will see that the apparent horizon is \emph{spacelike} for decelerating cosmologies.

The two examples we need for the main body of the paper are exponential inflation and power-law cosmologies. For exponential inflation one has $a(t) = a_0 \exp H_0 t$ with $H_0$ constant, and so $r_{AH}(t) = H_0^{-1} e^{-H_0 t}$. Note that in co-moving coordinates the horizon ``moves in'' toward $\Oi$, but the proper distance is $r_{prop} = a r_{AH} \equiv H_0^{-1}$. In this case the apparent horizon coincides with the event horizon. 

For power law cosmologies, we take $a(t) = a_0 (t/t_0)^{\alpha}$. One finds immediately that the co-moving radius of the apparent horizon is $r_{AH}(t) = \frac{1}{a_0 |\alpha|} t^{1-\alpha}$. In terms of the proper distance $r_{prop} = a r_{AH} = t/|\alpha|$, we see that the apparent horizon is always receding, and its proper distance will diverge to $\infty$ at late times regardless of the value of $\alpha$. However, it is also instructive to understand the picture in co-moving coordinates. We see that in terms of co-moving distance, the horizon is moving toward $\Oi$ if $\alpha > 1$, moving outward if $\alpha < 1$, and sitting at fixed coordinate if $\alpha = 1$; these are precisely the conditions for the scale factor to be accelerating, decelerating, or constant, respectively. 

Dynamics in cosmology is often parametrized by the apparent horizon. For example, when we compute the fluctuation spectrum of inflationary perturbations, the standard picture is that a field mode of wavenumber $k$ will ``freeze out'' when
\be
\frac{k}{aH} = k r_{AH} \ll 1
\ee
i.e. when the mode has a wavelength larger than the apparent horizon's radius, and one talks about modes ``exiting'' and ``entering'' the horizon when their wavelength crosses above or below this scale. In particular, things behind the apparent horizon (but within the event horizon) can re-enter the horizon during a phase of deceleration.

\bibliographystyle{JHEP} 
\bibliography{scrambling4}

\end{document}